\documentclass[journal]{IEEEtran}
\ifCLASSINFOpdf
\else
\fi

\usepackage{amsmath}
\usepackage{amssymb}
\usepackage{graphicx}
\usepackage{dcolumn}
\usepackage{bm}
\usepackage{mathrsfs}
\usepackage{epstopdf}
\usepackage{color}
\begin{document}
%
\title{Dissipative Quantum Electromagnetics}
%
%
%

\author{Wei~E.~I.~Sha,~\IEEEmembership{Senior Member,~IEEE,}
        Aiyin~Y.~Liu,
        Weng~Cho~Chew,~\IEEEmembership{Fellow,~IEEE}
\thanks{W. E. I. Sha is with Key Laboratory of Micro-nano Electronic Devices and Smart Systems of Zhejiang Province, College of Information Science \& Electronic Engineering, Zhejiang University, Hangzhou 310027, China (e-mail: weisha@zju.edu.cn).  A. Y. Liu is with the Department of Electrical and Computer Engineering, University of Illinois at Urbana-Champaign, Champaign, IL 61801 USA (e-mail: liu141@illinois.edu).  W. C. Chew is with Department of Electrical and Computer Engineering, Purdue University, West Lafayette, IN 47907, USA (e-mail: wcchew@purdue.edu).}
\thanks{Manuscript received Jan 29, 2018; revised XXX XXX, 2018.}}

%
%

\markboth{Journal of \LaTeX\ Class Files,~Vol.~XXX, No.~XXX, XXX~2018}%
{Shell \MakeLowercase{\textit{et al.}}: Bare Demo of IEEEtran.cls for IEEE Journals}
%



\maketitle
\newcommand{\hatb}[1]{{\hat{\bf #1}}}
\renewcommand{\vec}[1]{\mbox{${\bf #1}$}}
\renewcommand{\v}[1]{{{\bf #1}}}
\newcommand{\dyad}[1]{\mbox{$\overline{ \v{#1} }$}}
\newcommand{\romvec}[1] {\mbox{\boldmath $#1$}}
\newcommand{\lagr}{\mbox{\boldmath $\lambda$}}
\newcommand{\intsol}{\mbox{\boldmath $\phi$}}
\newcommand{\rhov}{\mbox{\boldmath $\rho$}}
\newcommand{\tr}{{\dagger}}
\newcommand{\pd}[2]{\frac{\partial #1}{\partial #2}}
\newcommand{\dd}[2]{\frac{\text{d} #1}{\text{d} #2}}
\newcommand{\od}[2]{\frac{d #1}{d #2}}

\newcommand{\funcd}[2]{\frac{\delta #1}{\delta #2}}

\newcommand{\commuteo}[2]{{\left[\hat #1,\hat #2\right]}}

\def\hatv #1{\hat{\v{#1}}}
\def\WRT{with respect to}
\def\IBP{integration by parts}
\def\CR{commutation relation}

\def\ED{\end{document}}

\def\dyad #1{\v{\overline{#1}}}
\def\dyadg #1{{\boldsymbol{\overline{#1}}}}
\def\vE{\v E}
\def\vH{\v H}
\def\vD{\v D}
\def\vB{\v B}


\newcommand{\vg} [1]{\mbox{\boldmath $#1$}}
\newcommand{\hatvg} [1]{\hat{\mbox{\boldmath $#1$}}}

\newcommand{\scriptl}{$\mathbb{L}$}

\def\be{\begin{equation} }
\def\ee{\end{equation} }

\def\domain #1{D(\mathbb{#1})}
\def\range #1{R(\mathbb{#1})}

\def\vromega{(\v r, \omega)}

\newcount\timehh\newcount\timemm
\timehh=\time \divide\timehh by 60 \timemm=\time \count255=\timehh
\multiply\count255 by -60 \advance\timemm by \count255
\def\draftheader{\slshape\today\ at
\ifnum\timehh<10 0\fi\number\timehh\,:\,\ifnum\timemm<10 0\fi\number\timemm}%

\def\norm #1{\left\| {#1} \right\|}

\def\Krylov#1{\mathcal K^{#1}\left(\dyad A,\v r_0\right)}

\newcommand{\jeq}{\mathbf J_{eq}}
\newcommand{\rbar}{(\mathbf r)}
\newcommand{\rbarp}{(\mathbf r')}
\newcommand{\bg}{\mathbf{\overline G}}
\newcommand{\prs}{(\mathbf r, \mathbf r')}
\newcommand{\meq}{\mathbf M_{eq}}
\newcommand{\hinc}{\mathbf H_{inc}}
\newcommand{\einc}{\mathbf E_{inc}}

\def\tdot{\hat t\cdot}
\def\Itd{\dyad I_t\cdot}
\def\It{\dyad I_t}

\newcommand{\js}{\mathbf J_{s}}
\newcommand{\bt}{\mathbf T}
\newcommand{\bbf}{\mathbf F}
\newcommand{\bmu}{\boldsymbol{\overline{\mu}}}
\newcommand{\beps}{\boldsymbol{\overline{\epsilon}}}
\newcommand{\bbtau}{\boldsymbol{\overline{\tau}}}
\newcommand{\bbI}{{\bf\overline{I}}}
\newcommand{\bbg}{\mathbf{\overline G}}
\newcommand{\bbi}{{\bf\overline I}}

\newcommand{\jays}{\mathbf J_s}
\newcommand{\ms}{\mathbf M_s}

\newcommand{\eneg}{e^{-ik_1z}}
\newcommand{\epos}{e^{ik_1z}}
\newcommand{\ein}{e^{in\phi}}

\def\Deltamur{\Delta\bmu_r}
\def\Deltaepsr{\Delta\beps_r}

\newcommand{\bbarz}{\mathbf{\overline Z}}
\newcommand{\bbarp}{\mathbf{\overline P}}
\newcommand{\bbark}{\mathbf{\overline K}}
\newcommand{\bbara}{\mathbf{\overline A}}
\newcommand{\bbarb}{\mathbf{\overline B}}
\newcommand{\bbard}{\mathbf{\overline D}}
\newcommand{\kayq}{{\mathbf{\overline K}}^{-1}\cdot\mathbf Q}
\def\MHE{electromagnetic equations}

\renewcommand{\v}[1]{{{\bf #1}}}
\def\dyad #1{\v{\overline{#1}}}
\def\braket#1#2{\langle{#1|#2}\rangle}
\def\bra#1{\langle{#1}|}
\def\ket#1{|{#1}\rangle}

\def\gTM{g^{\textrm{\scriptsize TM}}}
\def\gTE{g^{\textrm{\scriptsize TE}}}
\def\gTEs{g_s^{\textrm{\scriptsize TE}}}
\def\gTMr{g^{\textrm{\scriptsize TM}}\left(\v{r},\v{r}'\right)}
\def\gTEr{g^{\textrm{\scriptsize TE}}\left(\v{r},\v{r}'\right)}
\def\gTEsr{g_s^{\textrm{\scriptsize TE}}\left(\v{r},\v{r}'\right)}

\def\gTMS{g^{\textrm{\scriptsize TM},S}}
\def\gTES{g^{\textrm{\scriptsize TE},S}}
\def\gTEsS{g_s^{\textrm{\scriptsize TE},S}}
\def\gTMSr{g^{\textrm{\scriptsize TM},S}\left(\v{r},\v{r}'\right)}
\def\gTESr{g^{\textrm{\scriptsize TE},S}\left(\v{r},\v{r}'\right)}
\def\gTEsSr{g_s^{\textrm{\scriptsize TE},S}\left(\v{r},\v{r}'\right)}

\def\DG{\dyad G}
\def\curl{\nabla\times}
\def\curlp{\nabla'\times}
\def\div{\nabla\cdot}
\def\divp{\nabla'\cdot}
\def\grad{\nabla}
\def\gradp{\nabla'}
\def\curlDG{\curl\DG}

\def\TMS{{{\rm TM},S}}
\def\TES{{{\rm TE},S}}

\def\opL{\mathcal L}
\def\opK{\mathcal K}

\def\kmns{k_{mn}^2}
\def\pr{(\v r)}
\def\prp{(\v r')}
\def\prs{(\vec{r}, \vec{r}')}

\def\prt{(\v r, t)}
\def\prpt{(\v r', t)}

\def\prw{(\v r, \omega)}
\def\prpw{(\v r', \omega)}
\def\prwn{(\v r, \omega_n)}
\def\pw{(\omega)}

\def\rec2pi{\frac1{2\pi}}

\def\DGLM{dyadic Green's function for layered media}

\def\eeqa{\end{eqnarray}}
\renewcommand{\v}[1]{{{\bf #1}}}

\def\unitnormal{\hat{n}}

\def\MHE{electromagnetic equations}
\def\LVS{linear vector space}
\def\IPS{inner product space}
\def\CEM{computational electromagnetics}
\def\be{\begin{equation} }
\def\ee{\end{equation} }
\def\cm{classical mechanics}
\def\qm{quantum mechanics}
\def\EOM{equation of motion}
\def\EsOM{equations of motion}
\def\QM{Quantum mechanics}
\def\SE{Schr\"{o}dinger equation}
\def\Schr{Schr\"{o}dinger}
\def\tr{\text{tr}}
\def\Tr{\text{Tr}}
\def\EMS{effective mass Schr\"odinger}
\def\rep{representation}
\def\EM{electromagnetic}
\def\PBC{periodic boundary condition}
\def\LHS{left-hand side}
\def\RHS{right-hand side}
\def\HO{harmonic oscillator}
\def\DHO{double harmonic oscillator}
\def\RWA{rotating wave approximation}


%

\def\inf{\infty}
\def\intinf{\int\limits_{-\inf}^{\quad\inf}}
\def\intsinf{\int\limits_{0}^{\quad\inf}}
\def\iintinf{\iint\limits_{-\inf}^{\qquad\inf}}
\def\pd #1#2{\frac {\partial #1}{\partial #2}}
\def\krho{k_\rho}
\def\krhoo{k_{\rho 0}}
\def\hg{\hat g}
\def\today{\ifcase\month\or January\or February\or March\or April\or
May\or June\or July\or August\or September\or October\or November\or
December \fi\space\number\day, \number\year}
\def\inf{\infty}%
\def\dbarL{\bar{\bar {L}}}%
\def\degree{$\,^o$ }
\def\bE{{\bf E}}%
\def\tE{\tilde E}
\def\bJ{{\bf J}}%
\def\vr{{\bf r}}%
\def\dG{\bar{\bf G}_o (\vr ,\vr'')}%
\def\sg{g_o(\vr,\vr')}
\def\dbar#1{\bar{\bar{#1}}}%
\def\endeq{\eqno{\ENUM}}
\def\krho{k_\rho}
\def\~{\tilde}
\def\^{\hat}
\def\pd#1#2{\frac{\partial#1}{\partial#2}}

\def\R{\Re e}
\def\I{\Im m}
\def\eqn#1{\eqno{\text{#1}}}
\def\cal{\fam2 }
\def\lover#1{\buildrel \leftarrow \over {#1}}
\def\PVint{{-}\hskip -13.pt\int\limits}
\def\pvint{{-}\hskip -10pt\int\limits}
\def\m{\multline }
\def\em{\endmultline }

\def\itemitems{\hangindent2\parindent \textindent}
\def\items{\hangindent\parindent \textindent}

\def\MTT{{\it IEEE Trans. Microwave Theory Tech.}  MTT-}
\def\AP{{\it IEEE Trans. Antennas Propagat.}  AP-}
\def\GE{{\it IEEE Trans. Geosci. Remote Sensing}  GE-}
\def\RS{{\it Radio Sci.}  }
\def\EL{{\it Electron. Lett.}  }
\def\ritem{\noindent\hangindent=1.pc\hangafter=1}
\def\eitem#1{\par\noindent\hangindent\parindent\hangafter=1
    \hbox {\bf #1\enspace}\ignorespaces}
\def\q{\quad}
\def\qq{\qquad}
\def\qqq{\qq\q}
\def\qqqq{\qqq\q}

\def\three{\text{$\text{I}\text{I}\text{I}$}}
\def\tinf{\text{\it inf\,}}
\def\math#1{\vphantom{\left(#1\right)}#1}
\def\sqrts#1{ {\sqrt{#1}\,} }
\def\sgn{\,\text{\rm sgn}}

\def\bul{$\bullet$\hskip 1pc}
\def\nl{\vskip \baselineskip}
\def\bitem{\item{\bul}}
\def\subitem#1{\hbox{\hskip 2pc\vtop{\hsize 5.5in\parindent 0pt #1}
\hfil}}
\def\cl#1{\centerline{#1}}
\def\ub#1{\underbar{#1}}
\def\ul#1{\underline{#1}}
\def\Cal#1{{\cal #1}}
\def\Real{\Re e~}
\def\Imag{\Im m~}

\def\Div{\nabla\cdot}
\def\Curl{\nabla\times}
\def\Grad{\nabla}

\def\dot#1{\overset{\bm .}{#1}}
\def\ddot#1{\overset{\bm{..}}{#1}}

\begin{abstract}
The dissipative quantum electromagnetics is introduced in
a comprehensive manner as a field-matter-bath coupling problem.
First, the matter is described by a cluster of Lorentz oscillators.
Then the Maxwellian free field is coupled to the Lorentz oscillators
to describe a frequency dispersive medium.  The classical
Hamiltonian is derived for such a coupled system, using Lorenz gauge
and decoupled scalar and vector potential formulations. The classical
equations of motion are derivable from the Hamiltonian using
Hamilton equations. Then the Hamiltonian is quantized with all the
pertinent variables with the introduction of commutators between the
variables and their conjugate pairs.   The quantum equations of
motion can be derived using the quantum Hamilton equations.   It can
be shown that such a quantization scheme preserves the quantum
commutators introduced.   Then a noise bath consisting of simple
harmonic oscillators is introduced and coupled to the matter
consisting of Lorentz oscillators to induce quantum loss.  Langevin
source emerges naturally in such a procedure, and it can be shown
that the results are consistent with the fluctuation dissipation
theorem, and the quantization procedure of Welsch's group.  The
advantage of the present procedure is that no diagonalization of the
Hamiltonian is necessary to arrive at the quantum equations of
motion.  Finally, we apply the quantization scheme to model spontaneous 
emission of a two-level polarized atom placed above a dielectric cylinder 
that supports a bound state in the continuum.
\end{abstract}


%
\IEEEpeerreviewmaketitle

\section{Introduction}

Quantum dissipation is an interesting topic that has been studied by
many researchers \cite{Hopfield,Caldeira&Leggett}.  In the field of
quantum optics, it has been discussed in books
\cite{Tannoudji,Mandel&Wolf,Scully&Zubairy,HAUS,Loudon,Vogel&Welsch,TAN,Garrison&Chiao,Kira&Koch,Gerry&Knight,Fox,Walls&Milburn}.
and reported in many papers
\cite{Loudon1,Glauber,Huttner&Barnett,Milonni1,Grunner&Welsch,DungEtal,Ruppin,DungEtal1,Scheel&Buhmann,Suttorp1,Suttorp2,Philbin}.
Its importance is underscored by recent papers in the field
\cite{Hughes1,Hughes2}.  Since the manipulation of single photon is
occurring in microwave regime, it is appropriate to call this
emerging field quantum electromagnetics \cite[ref.
therein]{Astafiev}.

Since the total energy of the universe is conserved or a constant,
the Hamiltonian of the quantum system of the universe is a constant
of motion, implying energy conservation.  In the quantum
representation, the Hamiltonian, which becomes an operator, is a
Hermitian operator with real eigenvalues, implying that the
eigenfunctions of the system cannot decay with time.
 However, when the universe is partitioned into
sum of quantum systems, energy can be transferred between these
partitioned systems, giving rise to energy decay in one system and
energy gain in another system.

A popular way to consider loss or dissipation in a quantum system is
to couple it to a heat/noise bath, or a bath of oscillators
\cite{Caldeira&Leggett,Huttner&Barnett}, where
\cite{Huttner&Barnett} deals specially with electromagnetic system.
In principle, the total quantum system is still Hermitian if no
energy can escape from this coupled system. However, the heat bath
has infinite degrees of freedom; therefore, when energy is
transferred from the quantum system to the heat bath, the chance of
reversibility of the energy transfer is highly unlikely. This is
also the root cause for the increase of entropy in a thermodynamic
system.

In the spirit of the fluctuation dissipation theorem \cite{CallenWelton, Kubo, AgarwalI, AgarwalIII},
which describes a system, classical or quantum mechanical, in thermal equilibrium
with its environment or a heat bath, the energy lost to and fed back from
the environment balances. This was discussed extensively in the seminal work of Agarwal \cite{AgarwalI, AgarwalIII}. The feedback of energy from the environment to the system can be described by Langevin sources \cite{Langevin}.
Hence, in the system coupled to a heat bath (or a noise bath since the Langevin sources are random as
in random noise), energy does flow in the reverse direction as noise, namely,
from the heat bath to the quantum system \cite{CallenWelton}.  But this is different
from time reversibility.

Similar idea applied to electromagnetics has also been fervently
studied up to recent years
\cite{Loudon,Milonni1,Grunner&Welsch,DungEtal,Ruppin,DungEtal1,Vogel&Welsch,Scheel&Buhmann,Suttorp1,Suttorp2,Philbin}.
In many of these models, the quantum system of interest is often
subsumed by the noise bath, and becomes ``one of them". Huttner and
Barnett \cite{Huttner&Barnett} presented a canonical quantization
scheme for the electromagnetic fields in lossy and dispersive media
with Fano diagonalization \cite{Fano}. This scheme is based on the
Hopfield model of such media \cite{Hopfield} where the atomic or
molecular excitations are approximated by simple harmonic
oscillators.  The corresponding bilinear Hamiltonian is diagonalized
by ``Bogoliubov-like" transformation. In the first step, the
polarization field and the heat bath together form dressed-matter
operators. In the second step, the dressed-matter operators are
combined with photon/field operators to obtain the diagonal
Hamiltonian with polariton operators. Gruner and Welsch
\cite{Grunner&Welsch,DungEtal} started with the postulated polariton
Hamiltonian and verified the preservation of equal-time canonical
commutation relations in the quantization procedure, by using the
fluctuation dissipation theorem and by connecting the Langevin noise
sources to the polariton operators.

\begin{figure}[h]
\includegraphics[width = 0.45\textwidth]{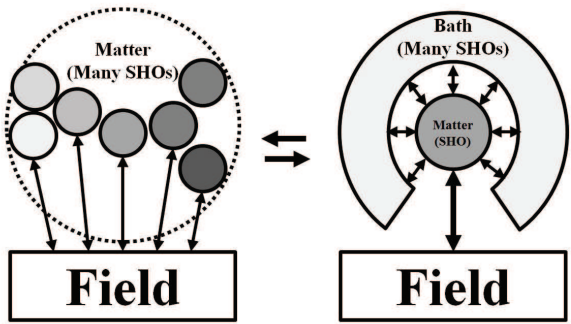}
\caption{\label{fig:1} Equivalence of the matter-field model with many species of matter oscillators \cite{Philbin} and the matter-field-bath model employed here.}
\end{figure}

Alternatively, based on the canonical quantization method proposed
by Glauber \cite{Glauber}, Milonni \cite{Milonni1} extended the mode
decomposition based quantization method to dispersive media by
employing the formulation of energy density in linear and dispersive
electromagnetic system \cite{Loudon} as the Hamiltonian density. Milonni
assumed that absorption is negligible, which approximates the
Kramers-Kronig relations \cite{Huttner&Barnett,Kramers&Kronig}.
In the work of Philbin \cite{Philbin}, more general electric and magnetic responses were incorporated by modeling the materials as a collection of harmonic oscillators with many frequencies. This work also offers an alternative view of the problem, in which the field-matter-bath system is reduced to a field-matter system. It is seen that as long as the matter system contains enough degrees of freedom its effect is the same as a bath, this is depicted in Figure \ref{fig:1}.

Consequently, a desired quantization approach for Maxwell's equations in lossy and
dispersive media requires: (1) an elegant Hamitonian with clear and
significant physical meaning; (2) a rigorous quantization procedure
satisfying canonical commutation relations; (3) an effective and
numerically-implementable susceptibility model satisfying
Kramers-Kronig relation.

Here, a lucid picture of quantum dissipation in electromagnetic
system is presented to fulfill the above requirements.
First, the field-matter system is quantized without mode
decomposition or diagonalization of the system.  Then the matter is
coupled to a noise bath to induce the appearance of dissipation and
Langevin sources. The choice to explicitly involve the bath is to maintain a strong connection with the classical system. To this end, we also use the quantization via the quantum Hamilton
equations \cite{QEM1,QEM2}. In this work, the matter will be
described by multi-species Lorentz oscillators, but only one species
will be considered here. It will be evident that the multi-species generalization is straightforward.

The following emphasizes the novelty of the approach presented here:
\begin{itemize}

\item The work starts with a Hamiltonian from first principles. Some of the previous
works started with Huttner and Barnett's result of Fano
diagonalization and polariton.
But here, diagonalization of the system of equations or mode
decomposition is not needed.

\item This approach starts with the time-domain Hamiltonian. Frequency domain methods follow naturally from time-domain ones if the system is linear.

\item Since a general heat-bath model is used, the heat-bath needs not
necessary be in thermal equilibrium as in fluctuation dissipation
theorem. Hence, it is possible to evolve this work along
non-equilibrium calculation as in non-equilibrium Green's function
approach adopted in quantum electron transport \cite{Keldysh,Datta}.

\item The high-loss model is different from that of Welsch's
group. It is able to deal with high loss in quantum electromagnetics
(QEM). The preservation of the commutation relations is rigorously
proved which leads to the same commutation relation of current as in
fluctuation dissipation theorem for both the low-loss and high-loss cases.
As shall be shown, the classical Drude-Lorentz-Sommerfeld (DLS) model is not suitable
for the high-loss case.

\item
Also, the commutation relation is shown to be satisfied in the
ensemble averaging sense after introducing phenomenological loss.
Furthermore, the Hamiltonian of the quantum system satisfies the
energy conservation in the ensemble average sense.

\end{itemize}

Before proceeding, we would like to address a large body of recent literatures that model quantum dissipation through a cascade of two-level systems \cite[ref. therein]{WelanderReview}. This approach is especially important for superconducting qubits, in which tunneling junction defects have been identified as the main cause of dissipation and decoherence. However, these defects still need to couple to phonon excitations of the junction material to cause dissipation. In such cases the oscillator model, though incomplete, is still important for gaining insights into the problem.

\section{Classical description of the field-matter system}

The interaction of electrons or charged particles of mass $m$ with
an electric field is often treated classically by the equation of
motion as
\begin{equation}\label{DLS:eq1a}
    m\ddot{x} + m\gamma\dot{x} + \kappa x = qE.
\end{equation}
On the left-hand-side, the first term is due to inertia since $\ddot
x$ is the acceleration of the charged particle.  The second term is
due to friction or collision since it is proportional to the
velocity $\dot x$ and $\gamma$ is the collision frequency. And the
third term is due to a restoring force similar to Hooke's law with the
spring constant $\kappa$.  The \RHS\ is the driving force due to the
electric field $E$ while $q$ is the particle charge. For lack of a
better name, this model will be called the DLS model
since these researchers have contributed to it at various times.
Sommerfeld has contributed to the quantum theory of such model.

Many models follow from this model.  When the friction term
dominates, the Drude model can be derived from it; when the inertia
term and the spring term dominate, the above is often called a
Lorentz oscillator;   also, when the spring term and the friction
term dominate, the Debye relaxation model follows. Furthermore,
when the collision and the spring terms are absent, the effective
permittivity for cold collisionless plasma can be derived. The above
equation can be easily converted into a form
\begin{equation}\label{DLS:eq1}
    \ddot{x} + \gamma\dot{x} + \omega_0^2 x = qE/m.
\end{equation}
where $\omega_0=\sqrt{\kappa/m}$ is the resonant frequency of the
oscillator if loss is absent or $\gamma=0$.
\textit{For the lossless case, the DLS oscillator becomes the Lorentz
oscillator which is just a simple harmonic oscillator}:  it can be
quantized.

\subsection{Coupling of Maxwellian Free Field to Lorentz Oscillators}


The quantization of free electromagnetic fields in a lossless and
dispersionless medium has recently been given using a different
viewpoint without mode decompositions \cite{QEM1,QEM2}.  In this
work, the field-matter-bath model will be introduced.  The ``field"
here refers to the free field or the photon field.  The ``matter"
here will be modeled by a collection of Lorentz oscillators as in
classical electromagnetics.  The ``bath" will be modeled by the
coupling of the Lorentz oscillators to an infinite random assortment
of simple harmonic oscillators which are also Lorentz oscillators.
This is similar to the Hopfield model and also similar to the
field-matter-bath model of Huttner and Barnett
\cite{Huttner&Barnett}.

However, using a similar approach here compared to before
\cite{QEM1,QEM2}, the quantization of free electromagnetic fields
coupled to lossless Lorentz dipoles is given without the need for
mode decomposition and diagonalization of the system. As shall be
shown, due to this coupling, the total field-matter system is
dispersive, but lossless. Next, loss can be induced by coupling the
field-matter system to a noise bath of harmonic oscillators, giving
rise to a field-matter-bath system with quantum dissipation.
Here, only one species of Lorentz dipoles is presented, but
generalization to multiple species of Lorentz dipoles can be easily
achieved so that arbitrary linear dispersive media can be modeled.
In Philbin's work \cite{Philbin}, an integral over the frequencies for these Lorentz oscillator fields are used to connect to the electric permittivity and magnetic susceptibilities of the linear dispersive materials.

%

To understand the dispersion better, the Maxwellian free field can
be thought of as a system where the dipoles are formed by the
polarization of electron-positron (e-p) pairs lurking in vacuum
\cite[p. 361]{Lancaster&Blundell}. The electric flux due to these
e-p pairs is given by $\v D=\varepsilon_0 \v E$, and their resonant
frequencies are so high that the $\varepsilon_0$ can be regarded as
dispersionless for the free field or the photon field.
In dispersive media consisting of field-matter coupling, the
Maxwellian system for the free or vacuum field is coupled to Lorentz
dipoles or oscillators representing the media. The Lorentz dipoles
model the oscillation of charged atoms or molecules that are bulky
and hence, have much larger inertial mass: They cannot be turned on
(or off) instantaneously, giving rise to dispersion.
Moreover, they have resonant frequencies much less than
those of the e-p pairs in vacuum \cite{Footnote1}.

\subsection{The Classical Equations for the Coupled System}

One starts with a single lossless Lorentz oscillator.   A distribution
of these Lorentz oscillators gives rise to polarization density, and
thus, polarization current.  A classical picture of this is given
in many text books.
%
%
For simplicity and without loss of generality, $\mu=\epsilon=1$ is
assumed \cite{Footnote2}. The polarization current can be easily derived from
\eqref{DLS:eq1}, and the pertinent equation can be written as
\begin{align}\label{hepv:eq1}
\ddot{ \v P}\prt +  \gamma \dot{ \v P}\prt +\omega_0^2 \v P\prt =
\omega_p^2  \v E\prt
\end{align}
where $\v P=nq\v x$, the plasma frequency is
$\omega_p^2=nq^2/(m\epsilon)$ where $n$ is the charge particle
density and $m$ is the mass of the charged particle with the charge of $q$.
Also, the normalized $\v E$ and $\v P$ have the same unit.
Consequently, Maxwell's equations can be augmented by coupling to
the polarization current as follows.
\begin{align}\label{hepv:eq2}
  \dot{\v H}\prt  &= -\Curl\v
E\prt\\\label{hepv:eq3}
 \dot{ \v E}\prt &=
\Curl\v H\prt -\v V\prt
\end{align}
where $\v P\prt$ is the polarization density, and $\v V\prt=\dot{ \v
P}\prt$ is the polarization current.  Here, $\v E$ and $\v H$ are
the free fields of the system.  The flux $\v D=\v E+\v P$ in the
present notation.
In addition, the above implies that $\Div\v H\prt=0$ and $\Div\v
E\prt=-\varrho_{\scriptscriptstyle P}\prt$,
%
%
where $\varrho_{\scriptscriptstyle P}\prt=\Div\v P\prt$, is the
polarization charge. Therefore, \eqref{hepv:eq1} can be rewritten as
two coupled first order equations in time \cite{Raman&Fan}
\begin{align}\label{hepv:eq4}
 \dot{\v P}\prt &= \v V\prt
\\\label{hepv:eq5}
 \dot{ \v V}\prt &= \omega_p^2 \v
 E\prt - \gamma \v V\prt
-\omega_0^2 \v P\prt
\end{align}
Hence, the loss in the system is represented through $\gamma\ne 0$.
In order to quantize the system of equations, $\gamma$ is set to
zero first to model the lossless case.  This gives rise to a
lossless Hermitian system that can be quantized similar to before
\cite{QEM1,QEM2}, albeit with some complication since the medium is
now dispersive.
Eventually, the system will be coupled to a bath of harmonic
oscillators, from which quantum dissipation follows.

\subsection{Lorenz Gauge and the Decoupled Potentials}


As mentioned before, the lossless dispersive case will be considered
first, as it is easily quantized due to the
Hermitian nature of the system. To this end, one needs to derive the
Hamiltonians that describe the above system. The Lorenz gauge will
be used here \cite{ChewPIER}, since it is commensurate with the special
relativity where space and time are treated on the same footing.
Therefore,
\begin{align}\label{lgdp:eq1}
\v E = -\dot{ \v A}-\nabla \Phi,\qquad \v H= \v B=\Curl\v A, \qquad
\Div \v A=-\dot{ \Phi}
\end{align}
Since $\mu=1$, for the free field here, $\v B=\v H$.
With these notations, one can show easily that the original \EsOM\
 for the
decoupled potential equations, derivable from the above are:
\begin{align}\label{lgdp:eq2}
\nabla^2\Phi\prt &-\ddot{ \Phi}\prt = \varrho_{\scriptscriptstyle P}\prt\\
 \Curl\Curl \v A\prt - \nabla&\Div \v A\prt + \ddot {\v
A}\prt = \v V\prt\label{lgdp:eq3}
\end{align}
The \EOM\ for the lossless Lorentz oscillator is then
\begin{align}\label{lgdp:eq4}
\dot {\v V}\prt +\omega_0^2 \v P\prt = \omega_p^2 \v E\prt
\end{align}
In the above, one assumes that $\Div \v P=\varrho_{\scriptscriptstyle P}$, and hence,
$\Div\v E=-\varrho_{\scriptscriptstyle P}$ which explains the positive sign of
$\varrho_{\scriptscriptstyle P}$ on the \RHS\ of \eqref{lgdp:eq2}.

\subsection{Derivation of the Classical Hamiltonian}

In order to quantize the electromagnetic system, its classical
Hamiltonian needs to be derived first.
%
%
%
To derive the corresponding classical Hamiltonians for the different
fields, conjugate momenta for $\v A$, $\Phi$, and $\v P$ are defined
as
\begin{align}\label{dch:eq1}
&\vg\Pi_A\prt=\dot{ \v A}\prt,\qquad\Pi_\Phi\prt= \dot
{\Phi}\prt,\nonumber\\& \vg \Pi_P=\beta \v V=\beta\dot{ \v P}\prt
\end{align}
Then similar to the method outlined in \cite{QEM1,QEM2}, the sources
on the \RHS\ of \eqref{lgdp:eq2} to \eqref{lgdp:eq4} can be treated
as impressed sources first. Then
 one arrives
at the Hamiltonian densities for the scalar potential, vector
potential, and the polarization density. They are:
\begin{align}
\mathscr H_A&= \frac12\left[\vg\Pi_A^2 + \left(\Curl\v
A\right)^2+\left(\Div \v A\right)^2-2\v A\cdot \v
V\right]\label{dch:eq2}\\
\mathscr H_\Phi &= \frac12\left[\Pi_\Phi^2 +
\left(\nabla \Phi\right)^2 + 2\Phi\varrho_{\scriptscriptstyle P}\right]\label{dch:eq3}\\
\mathscr H_P
&=\frac12\left[\vg \Pi_P^2/\beta + f \v P^2-2\v P\cdot\v
E\right]\label{dch:eq4}
\end{align}
where $\beta=1/\omega_p^2$, $f=\omega_0^2/\omega_p^2$. The
Hamiltonian is related to the Hamiltonian density via
\begin{equation}\label{dch:eq5}
H = \int\!dx^4 (\mathscr H_A-\mathscr H_\Phi+\mathscr H_P)
\end{equation}
In the above, $dx^4=dtd\vr$;  integration over four space (space and time) is
necessary since in a dispersive medium, the fields from different
times affect each other.  It is also implicitly implied that the
above Hamiltonian densities are integrated over four space that
allows the invocation of integration by parts.

It is to be noted that in \eqref{dch:eq2} to \eqref{dch:eq4}, there
exist coupling between the different Hamiltonians. The Hamiltonian
for $\v A$ has the polarization current $\v V$ in it, while that for
$\Phi$ has the polarization charge $\varrho_{\scriptscriptstyle P}$
in it, and that for $\v P$ has $\v E$ in it, where $\v E$ is related
to $\v A$ and $\Phi$ via \eqref{lgdp:eq1}.  But so far, the sources
on the \RHS\ of \eqref{lgdp:eq2} to \eqref{lgdp:eq4} are treated as
``impressed sources" as expounded in \cite{QEM2}. This is incorrect
as ``impressed sources" cannot appear if we are writing down a Hamiltonian for the entire system, with no external influence.

To remedy this, the total Hamiltonian is the sum of the three
Hamiltonians plus the interaction energy. As shall be seen, it is
via the interaction energy that the coupling occurs.   Hence, the
corrected Hamiltonian becomes
\begin{align}\label{dch:eq6}
H = \int\!dx^4 (\mathscr H_A- \mathscr H_\Phi + \mathscr H_P + 2\v
E\cdot\v P- \Phi\varrho_{\scriptscriptstyle P})\nonumber\\=\int\! dx^4 (\mathscr H_A- \mathscr H_\Phi
+ \mathscr H_P + 2\v A\cdot\v V + \Phi\varrho_{\scriptscriptstyle P})
\end{align}
The above equality can be shown using integration by parts over
space and time.
It can be further shown, using integration by parts in space and
time, that the $\v A\cdot \v V$ cancels a term in $\v P\cdot \v E$,
leaving behind a $\v P\cdot\nabla\Phi$ term in $\mathscr H_P$. The
sum Hamiltonian then becomes
\begin{align}\label{dch:eq7}
H =& \int\!dx^4\,\frac12\left[\vg\Pi_A^2 + \left(\Curl \v
A\right)^2+\left(\Div \v A\right)^2 \right.\nonumber\\&-\left. \Pi_\Phi^2 - \left(\nabla
\Phi\right)^2 + \vg \Pi_P^2/\beta + f \v P^2 + 2\v P\cdot\nabla\Phi
\right]
\end{align}
It can now be shown that the above is in fact (see Appendix
\ref{app:1})
\begin{align}\label{dch:eq8}
H = \int \!dx^4\,\frac12\left[\v E^2+\v H^2 + \beta\v V^2 + f \v P^2
\right]
\end{align}
The above result is physically important because the first two terms
correspond to energy stored in the electric and magnetic free
fields, respectively; the third and the fourth terms are the kinetic
and potential energies stored in the Lorentz oscillator,
respectively. The above result is comforting since it implies that
the Hamilton is equal to the total energy of the system: This has to
be a constant of motion.

However, the \EsOM\ in \eqref{lgdp:eq2} to
\eqref{lgdp:eq4} cannot be derived readily from \eqref{dch:eq7}. The
reason is that as the free field is produced by the Lorentz
oscillator, there is a back action of the free field onto the
Lorentz oscillator. The Hamiltonians used in equations
\eqref{dch:eq2}, \eqref{dch:eq3}, and \eqref{dch:eq4} do not
consider this back action. Therefore, when this back action is
present, the conjugate momentum of $\v A$ has to be redefined.


To this end, a new conjugate momentum for $\v A$, namely
$\vg\Pi_{AP}=\dot{\v A}-\v P$, is defined.  Then \eqref{dch:eq7}
becomes \cite{Footnote4}
\begin{align}\label{dch:eq11}
H = &\int\!d\v r\,\frac12\left[\left(\vg\Pi_{AP}+\v P\right)^2 +
\left(\Curl \v A\right)^2+\left(\Div \v A\right)^2 \right.\nonumber\\&-\left. \Pi_\Phi^2 -
\left(\nabla \Phi\right)^2 + \vg \Pi_P^2/\beta + f \v P^2 + 2\v
P\cdot\nabla\Phi
\right]
\end{align}
In the above, the Hamiltonian involves now an integral over
three-space since this Hamiltonian remains to be a constant of
motion for an energy conserving system.
Hamilton equations, similar to those given in \cite{QEM1,QEM2}, can
now be invoked to derive the \EsOM\ for the conjugate variables,
namely,
\begin{align}\label{dch:eq12}
\dot{\v A}(\v r,t)&=   \funcd{H}{ \vg\Pi_{AP}(\v r,t)},&
\dot{\vg\Pi}_{AP}(\v r,t)&= -\funcd{H}{ \v A(\v r,t)}
\\\label{dch:eq13}
\dot{\Phi}(\v r,t)&=  -\funcd{H}{ \Pi_\Phi(\v r,t)},&
\dot{\Pi}_\Phi(\v r,t)&= \funcd{H}{ \Phi(\v r,t)}
\\\label{dch:eq14}
\dot{\v P}(\v r,t)&=   \funcd{H}{ \vg \Pi_P(\v r,t)},& \dot{\vg
\Pi}_P
 (\v r,t) &= -\funcd{H}{ \v P(\v r,t)}
\end{align}
The minus sign found in the equations of motion for $\Phi$ and
$\Pi_\Phi$ is because the total Hamiltonian is proportional to
Hamiltonian for the vector potential minus the Hamiltonian for the
scalar potential.  This was further explained in \cite{QEM1,QEM2}.

It can be shown easily that the \EsOM\, \eqref{lgdp:eq2},
\eqref{lgdp:eq3}, \eqref{lgdp:eq4}, can be re-derived from the
above. The above gives the classical Hamiltonian description of the
system. Once the classical Hamiltonian description of the system is
available, the quantum Hamiltonian description of the system can be
arrived at.  But since such exercise is infrequent, the procedure
will be further elaborated next.  The left column above corresponds
to the \EsOM\ for $\v A$, $\Phi$ and $\v P$.  They correspond to
taking the variation of the Hamiltonian with respect to the
conjugate momenta $\vg\Pi_{AP}$, $\Pi_\Phi$, and $\v V$, and they
can be easily done. Therefore, evaluating the \RHS s of all the above
equations by taking the proper functional derivatives of the
Hamiltonian, one arrives at 6 equations
\begin{align}
\label{dch:eq15}
\dot{\v A}(\v r,t)&=\vg\Pi_{AP}\prt+\v P\prt, \nonumber\\
\dot{\vg\Pi}_{AP}(\v r,t)&= -\Curl\Curl \v A\prt + \nabla\Div \v A\prt
\end{align}
\vspace*{-\baselineskip}
\begin{align}
\label{dch:eq16}
\dot{\Phi}(\v r,t)&=\Pi_\Phi\prt\nonumber\\
\dot{\Pi}_\Phi(\v r,t)&=\nabla^2\Phi - \varrho_{\scriptscriptstyle P}
\end{align}
\vspace*{-\baselineskip}
\begin{align}
\label{dch:eq17}
\dot{\v P}(\v r,t)&= \vg\Pi_P\prt/\beta=\v V\prt, \nonumber\\
\beta\dot{\v V}(\v r,t) &= -f\v P\prt-\vg \Pi_{AP}\prt - \nabla\Phi\prt-\v P\prt\notag\\
&=-f\v P\prt + \v E\prt
\end{align}
The equations above contain the equations of motion for the original
variables, $\v A$, $\Phi$, and $\v P$, and the conjugate variables
$\vg\Pi_{AP}$, $\Pi_\Phi$, and $\v V$. They can be combined to yield
the classical equations of motion in \eqref{lgdp:eq2} to
\eqref{lgdp:eq4}.
It is to be reminded that when functional derivatives in
\eqref{dch:eq12} to \eqref{dch:eq14} are taken, they yield Dirac
delta functions with the sifting property similar to the rule expounded
in \cite{QEM1}.  Examples of which are
\begin{align}\label{cho:eq26}
\funcd{\v A(\v r,t)}{ \v A(\v r',t)}=\funcd{\vg \Pi (\v r,t)}{ \vg
\Pi(\v r',t)}=\dyad I\delta(\v r-\v r'),\\\funcd{\v A(\v r,t)}{
\vg\Pi (\v r',t)}=\funcd{ \vg\Pi (\v r,t)}{ \v A(\v r',t)}=0
\end{align}
Their use greatly simplifies the integral of the Hamiltonian.

It is pleasing to note that the \EsOM\ of the total field-matter
system: Maxwellian free fields coupled to the Lorentz oscillators
can be expressed in terms of Hamilton \EsOM.  This is definitely a
more complicated system than that for the Maxwellian free fields
alone.  Although the Coulomb gauge which quantizes
only the transverse fields is widely used in quantum optics,
we would like to present a different quantization approach under the Lorenz gauge in this work.
In view of gauge invariance and the numerical niceties of the Lorenz gauge,
we feel that this scheme should be considered as an alternative
which may present more choices when integrating the capabilities
of computational electromagnetics with quantum optics \cite{Explanation}.


\section{Quantum description of the field-matter system}


To arrive at the quantum equations of motion, first, the quantum
Hamiltonian corresponding to the above has to be derived. Hence, to
arrive at the quantum equivalence of the above classical system, it
is necessary to first elevate all the conjugate variables to become
quantum operators.   With this, the Hamiltonian becomes an operator
as well, and is now governed by
\begin{align}\label{qef:eq1}
\hat H = &\int\!d{\bf r}\,\frac12\left[\left(\hat{\vg\Pi}_{AP}+\hat{\v
P}\right)^2 + \left(\Curl\hat{ \v A}\right)^2+\left(\Div \hat{\v
A}\right)^2 \right. \nonumber\\ & \left.- \hat{\Pi}_\Phi^2 - \left(\nabla \hat{\Phi}\right)^2 +
\hat{\vg \Pi}_P^2/\beta + f\hat{\v P}^2 + 2\hat{\v
P}\cdot\nabla\hat{\Phi} \right]
\end{align}
These quantum operators corresponding to the conjugate variables
operate on a quantum state $|\psi\rangle$ and the time evolution of
the entire quantum system described by $\hat H$ is now given
by \cite{Footnote3}
\begin{equation}\label{qef:eq2}
\hat H |\psi\rangle = i\hbar \partial_t |\psi\rangle
\end{equation}
The time evolution of each operator in this quantum system is given
by the Heisenberg equation of motion:
\begin{equation}\label{qef:eq3}
i\hbar\dot{ \hat O} = \left[\hat O,\hat H \right]
\end{equation}
It turns out that the quantum Hamilton equations can be derived from
the above \cite{QEM1,QEM2,Louisell}.  To this end, it is necessary that one
defines the commutation relations between operators that are
conjugate to each other.  They are important for energy
conservation.  Hence, the commutation relations that should be
introduced here are: \cite{Footnote5}
\begin{align}\label{qef:eq4}
\commuteo{{\v \Pi}_{AP}\prt}{{\v A}\prpt} = -i\hbar \delta(\v r-\v
r') \hat{\dyad I}\\\label{qef:eq5}
\commuteo{\Pi_\Phi\prt}{\Phi\prpt}=+i\hbar\delta(\v r-\v r')\hat I
\\\label{qef:eq6} \commuteo{{\vg \Pi}_P\prt}{{\v P}\prpt} = -i \hbar \delta(\v
r-\v r') \hat{\dyad I}
\end{align}
Since the variables $\v A$, $\Phi$, and $\v P$ are independent
variables so are their conjugate variables, when elevated to be
quantum operators, they are also mutually commuting. The above
commutators for the fields are similar and analogous in spirit to
the position-momentum commutator
\begin{align}\label{qef:eq7}
\commuteo{q_i(t)}{p_j(t)}=i\hbar\hat I \delta_{ij}
\end{align}   The above
commutator in \eqref{qef:eq7} induces the derivative operators that
can be used to abbreviate the Heisenberg equation of motion as shown
previously \cite{QEM1,QEM2,Louisell}. Namely, in the discrete case, they are
\begin{align}\label{qef:eq8}
\commuteo{p_{i'}}{q_i^n} &=
 -i n\delta_{ii'}\hat q_i^{n-1}\hbar =-i\hbar \pd{}{\hat q_{i'}} \hat
 q_i^n\nonumber\\
 \commuteo{p_{i'}}{ H} &=-i\hbar \pd{}{\hat
 q_{i'}} \hat H=i\hbar\dot{\hat p}_{i'}(t) \\
 \label{qef:eq9}
\commuteo{q_{i'}}{p_i^n} &=
 i n\delta_{ii'}\hat p_i^{n-1}\hbar =i\hbar \pd{}{\hat p_{i'}} \hat
 p_i^n\nonumber\\
 \commuteo{q_{i'}}{ H} &=i\hbar \pd{}{\hat
 p_{i'}} \hat H=i\hbar\dot{\hat q}_{i'}(t)
\end{align}
For the continuum case, the above commutators, \eqref{qef:eq4} to
\eqref{qef:eq6} induce functional derivative operations
\cite{QEM1,QEM2}. Consequently, the quantum Hamilton equations of
motion are
\begin{align}\label{qef:eq10}
\commuteo{{\vg\Pi}_{AP}\prt}{H} &=-i\hbar\,\funcd{{\hat H}}{\hat
 {\v A}\prt}
=i\hbar\dot{\hat {\vg\Pi}}_{AP}\prt
\\\label{qef:eq11}
\commuteo{{\v A}\prt}{ H}&=i\hbar\,\funcd{\hat H}{\hat
 {\vg\Pi}_{AP} \prt} =i\hbar\,\dot{\hat {\v A}}\prt\\\label{qef:eq11a}
\commuteo{\Pi_\Phi\prt}{H}&=i\hbar\,\funcd{\hat H}{{\hat
 \Phi}\prt} =i\hbar\,\dot{\hat \Pi}_\Phi\prt
\\\label{qef:eq12}
\commuteo{\Phi\prt}{ H }&=-i\hbar\,\funcd{\hat H }{\hat
 \Pi_\Phi\prt} =i\hbar\dot{\hat \Phi}\prt\\
\commuteo{{\vg \Pi}_P\prt}{H} &=-i{\hbar}\,\funcd{{\hat H}}{\hat
 {\v P}\prt}
=i\hbar\,\dot{\hat {\v \Pi}}_P\prt
\\\label{qef:eq13}
\commuteo{{\v P}\prt}{ H }&=i{\hbar}\,\funcd{\hat H }{\hat
 {\vg \Pi}_P \prt} =i\hbar\,\dot{\hat {\v P}}\prt
\end{align}
The above quantum Hamilton equations are very similar to their
classical counterparts in \eqref{dch:eq12} to \eqref{dch:eq14}, and
then in \eqref{dch:eq15} to \eqref{dch:eq17}. Hence, the quantum
analogues of \eqref{lgdp:eq2}-\eqref{lgdp:eq4} can be obtained as in
previous work \cite{QEM1,QEM2}.
From them, the \EsOM\ of the quantum operators that are the analog
of \eqref{hepv:eq1} to \eqref{hepv:eq3} for the lossless case are
\begin{align}\label{qef:eq14}
\ddot{ \hat{\v P}}\prt   &+\omega_0^2\hat{ \v
P}\prt = \omega_p^2  \hat{\v E}\prt\\
\label{qef:eq15}
 \dot {\hat{\v H}}\prt  &= -\Curl\hat{\v
E}\prt\\\label{qef:eq16}
 \dot{ \hat{\v E}}\prt  &=
\Curl\hat{\v H}\prt -\hat{\v V}\prt
\end{align}
It is to be noted that the above \EsOM\ of these quantum operators
have meaning only if they operate on a quantum state $|\psi\rangle$
of the system.  Also, the above quantum equations are derived
without the normal mode decomposition approach, but can be derived
directly from the Hamiltonian using the quantum Hamilton equations.
Hence, diagonalization of the system is not necessary.
Normal mode decomposition is possible for all linear systems in
theory, but for some practical systems, they have to be done
numerically.   This quantization approach here avoids having to find
the normal modes of the system.
%

\subsection{Preservation of Commutators}

It can be shown that the commutators in \eqref{qef:eq4} to
\eqref{qef:eq6} are still preserved with the above quantization
procedure.  For instance, for the commutator
\begin{align}
\hat {\v C}_A =\left[{{\hat{\vg \Pi}_{AP}}},{\hat{\v A}}\right]
\end{align}
then,
\begin{align}
\frac{d}{dt} \hat {\v C}_A &= \left[{\dot{\hat{ \vg
\Pi}}_{AP}},\hat{\v A}\right] + \left[{ \hat{\vg \Pi}_{AP}},{\dot{\hat{\v A}}}\right]\notag\\
&=\left[-\Curl\Curl\hat{\v A}+\nabla\nabla\cdot \hat{\v A} ,\hat{\v
A}\right] + \left[\dot{\hat{\v A}}-\hat{\v P},\dot{\hat{\v A}}
\right]\nonumber\\&=0
\end{align}
The \RHS\ is zero because it can be shown that by using the discrete
version of a field as expounded in \cite{QEM1}, the space derivative
of a field operator commutes with the field operator itself.  Also,
$\hat{\v P}$ and $\dot{\hat{\v A}}$ commute because $\v P$ and
$\dot{\v A}$ are independent variables.  Similar method can be used
to show that the rest of the commutators in \eqref{qef:eq4} to
\eqref{qef:eq6} are preserved in this quantization scheme. The
preservation of commutators is related to the conservation of
energy.

\section{Dissipation by Coupling to a Noise Bath}


The topic of quantum dissipation in quantum optics has been reported
in many books
\cite{Tannoudji,Mandel&Wolf,Scully&Zubairy,HAUS,Loudon,Vogel&Welsch,TAN,Garrison&Chiao,Kira&Koch,Gerry&Knight,Fox,Walls&Milburn}.
In this section, the coupling of a single Lorentz harmonic oscillator
to a noise bath will be demonstrated.
Both classical and quantum dissipations can be induced by coupling
the non-dissipative system to a noise bath of harmonic oscillators
\cite{Caldeira&Leggett,Huttner&Barnett}.  In classical problems, the
loss in the DLS oscillator is due to the collisions
of electrons with the lattice or the ions. Hence, it is
reasonable to assume that the loss introduced to the Lorentz oscillator comes
from its coupling to other systems which can be modeled as a noise
bath.

In general, there are many sources of noise in a system.  To
simplify, the noise bath will be assumed to consist only of a large
collection of simple harmonic oscillators. This model has been
assumed in the Huttner and Barnett model \cite{Huttner&Barnett} as
well as in other quantum systems
\cite{Caldeira&Leggett,Suttorp1,Philbin}. There in
\cite{Huttner&Barnett}, the free-field is coupled to matter, and
then the matter is coupled to a noise bath. The matter there is
equivalent to the Lorentz oscillator here.  Because of the
simplification of the noise bath model, the noise bath is only
modeled phenomenologically.



Hence, in this work, it is assumed that the loss in the DLS oscillator
is a consequence of coupling of a Lorentz oscillator to a noise bath which
is modeled by a collection of harmonic oscillators.
%
%
%

\subsection{Classical Case}

First, the classical Hamiltonian due to the coupling of the
field-matter to a noise bath will be illustrated.  To this end, the
total Hamiltonian density is given by
\begin{align}\label{DCNB:eq1}
\mathscr{ H}_{PB}=\mathscr{ H}_{P}+\mathscr{ H}_{B}+\mathscr{
H}_{INT}
\end{align}
where on the \RHS, the first Hamiltonian density, $\mathscr{
H}_{P}$, is due to matter consisting of Lorentz oscillators which
are used to describe the polarization current, the second
Hamiltonian density, $\mathscr{ H}_{B}$, is due to the noise bath,
while the third Hamiltonian density is the interaction between the
matter and the bath.  The Hamiltonian density $\mathscr{ H}_{P}$ is
as before, and reproduced here as:
\begin{align}\label{DCNB:eq2}
\mathscr{ H}_P = \frac12\left[ \beta{\v V}^2 + f{\v P}^2 +2{\v
P}\cdot{\v E}
 \right]
\end{align}
 The bath Hamiltonian consisting of a large random assortment of simple harmonic oscillators  can be
 written as \cite{Footnote7}
\begin{align}\label{DCNB:eq3}
\mathscr{ H}_{B}=\sum_{j}\frac12\left[\vg \Pi_{P,j}^2/\beta_j + f_j
\v P_j^2\right]
\end{align}
where $\vg \Pi_{P,j}=\beta_j\v V_j$ is the conjugate variable to $\v
P_j$. The interaction Hamiltonian is
\begin{align}\label{DCNB:eq4}
\mathscr{ H}_{INT}=\sum_{j}\left[ \alpha_j^\Pi\vg \Pi_{P,j}\cdot\vg
\Pi_P + \alpha_j^P \v P_j\cdot \v P \right]
\end{align}
The above is motivated by the discrete case: If one has $N$ discrete
oscillators, coupled by a stiffness matrix $K_{ij}$, the coupling
term yielding the potential energy is proportional to $\sum_{i,j}
q_i K_{ij} q_j$. Similar argument can be made for coupling via the
mass matrix \cite{QEM1,Goldstein}.

The matter-bath Hamiltonian $H_{PB}$ can be obtained by integrating
the Hamiltonian density in \eqref{DCNB:eq1} over space.
The \EsOM\ can then be derived such that
\begin{align}
\label{DCNB:eq5}
\dot{\v P}(\v r,t)= \funcd{H_{PB}}{ \vg \Pi_P(\v r,t)}= &\v V\prt+\sum_j \alpha_j^\Pi\vg\Pi_{P,j}\prt\\
\label{DCNB:eq6}
\dot{\vg \Pi}_P(\v r,t) = -\funcd{H_{PB}}{ \v P(\v r,t)}= &-f\v P\prt+\v E\prt \nonumber\\
& -\sum_j\alpha_j^P \v P_j\prt
\\
\label{DCNB:eq7}
\dot{\v P}_j(\v r,t)= \funcd{H_{PB}}{ \vg \Pi_{P,j}(\v r,t)}= &\v V_j\prt+\alpha_j^\Pi\vg{\Pi}_P\prt\\
\label{DCNB:eq8}
\dot{\vg \Pi}_{P,j}(\v r,t) = -\funcd{H_{PB}}{ \v P_j(\v r,t)}= &-f_j\v P_j\prt -\alpha_j^P \v P\prt
\end{align}
 It is to be noted that in the last equation above, the \EOM\ for
 the noise bath oscillator is only coupled to the polarization current $\v
 P\prt$.  This is unlike \eqref{DCNB:eq6} above or \eqref{dch:eq17},
 where the \EOM\ for the polarization current is coupled to the
 driving field $\v E\prt$.

\subsection{Quantum Case}

The derivation of the \EsOM\ for the quantum case is quite routine:
First, the conjugate variable pairs are elevated to be operators,
and then commutators between them are defined. The new commutator
needed here for the bath oscillators is
\begin{align}\label{DCNB:eq9}
 \commuteo{{\vg \Pi}_{P,j}\prt}{{\v P}_{j'}\prpt} = -i \hbar
\delta(\v r-\v r')\delta_{jj'} \hat{\dyad I}
\end{align}
The relevant Hamiltonian is hence elevated to be an operator. Using
the quantum Hamilton equations, the \EsOM\ can be derived for the
matter-bath coupling to be
\begin{equation}
\dot{\hat{\v P}}(\v r,t)= \funcd{\hat H}{ \hat{\vg \Pi}_P(\v r,t)}=\hat{\v V}\prt+\sum_j \alpha_j^\Pi\hat{\vg\Pi}_{P,j}\prt
\end{equation}
\begin{align}
\dot{\hat{\vg \Pi}}_P(\v r,t) = -\funcd{\hat H}{ \hat{\v P}(\v r,t)}= &-f\hat{\v P}\prt+\hat{\v E}\prt\nonumber\\
&-\sum_j\alpha_j^P\hat{ \v P}_j\prt
\end{align}
\begin{equation}
\dot{\hat{\v P}}_j(\v r,t)= \funcd{\hat H}{\hat{ \vg \Pi}_{P,j}(\v r,t)}=\hat{\v V}_j\prt+\alpha_j^\Pi\hat{\vg{\Pi}}_P\prt
\end{equation}
\begin{equation}
\dot{\hat{\vg \Pi}}_{P,j}(\v r,t) = -\funcd{\hat H}{ \hat{\v P}_j(\v r,t)}=-f_j\hat{\v P}_j\prt-\alpha_j^P \hat{\v P}\prt
 \end{equation}
Similar to before, it is easy to show that these commutators in
\eqref{qef:eq6} and \eqref{DCNB:eq9} are preserved by the above
\EsOM.


\subsection{Asymptotic Solution--Coupling of the Lorentz Oscillator to a Noise Bath}

The above coupled \EsOM\ has no closed form or analytic solution.
But in the limit when the bath is assumed to be infinitely large,
approximate analytic solution can be obtained.  Before this is
shown, it is necessary to simplify the above \EsOM.   A closer look
at the \EsOM\ shows that they are entirely local: Namely,  the
Lorentz oscillators are not mutually coupled to each others, unlike
the e-p pair oscillators in Maxwell's equations.  Moreover, the
($x$, $y$, $z$) components of the oscillations are entirely
independent of each other. Therefore, they can be fully described by
scalar oscillators at a given location.  This is also the spirit of
the matter-bath model in works of other researchers
\cite{Huttner&Barnett,Suttorp1,Philbin}.

Consequently, the matter-bath model can be understood by studying only one
single oscillator's coupling to a noise bath.  (Only the driving
field $\v E\prt$ is a function of position.)  To this end, the above
problem will be reduced to one involving a single harmonic oscillator
coupled to a noise bath.  Therefore, the following replacements are
 assumed next:
\begin{align}\label{qd:eq2}
\hbar\omega_0\hat \zeta^2 \leftrightarrow f\hat{\v P}^2,\qquad
\hbar\omega_0\hat \pi^2 \leftrightarrow \beta\hat{\v V}^2,\nonumber\\
\hbar\omega_j\hat \zeta_j^2 \leftrightarrow f_j\hat{\v P}_j^2,\qquad
\hbar\omega_j\hat \pi_j^2 \leftrightarrow \beta_j\hat{\v V}_j^2
\end{align}
Then the Hamiltonian for a single Lorentz oscillator coupled to a
noise bath becomes
%
%
\begin{align}\label{qd:eq4}
{\hat H}_{PB} = &\frac12\hbar\omega_0 \left( \hat{\pi}^2 +
\hat{\zeta}^2\right) + \frac12\hbar\sum_j\omega_j \left(
\hat{\pi_j}^2 + \hat{\zeta_j}^2\right)\nonumber\\
&+\sum_j \hbar\left(\alpha_j^\zeta \hat\zeta\hat\zeta_j+\alpha_j^\pi
\hat\pi\hat\pi_j\right)+2C\hat{\zeta}\hat{ E}
\end{align}
The first term represents the Hamiltonian for the single Lorentz
oscillator yielding the polarization current inside the medium,
while the second term represents the Hamiltonian of the harmonic
oscillators in the bath.  The third term is the interaction of the
single Lorentz oscillator with the bath oscillators.  The last term
can be ignored for the following analyses. Because the matter is coupled
to the noise bath only and $E$ can be regarded as the external driving field.
In this model, the Lorentz oscillator and the
noise oscillators at different locations are completely independent
of each other, hence, $\omega_0$, $\omega_j$, $\alpha_j^\zeta$,
$\alpha_j^\pi$, $C$, and $\hat E$ can be functions of positions.  In
\eqref{qd:eq2}, these equations are similar to those of the mode
decomposition picture \cite{QEM2}, but they represent modes at
different locations.

Without the last term, the above Hamiltonian is that of a collection
of $N$ coupled simple harmonic oscillators.  They will have $N$
natural modes or resonant frequencies.  The number of modes will
become infinitely large as $N\rightarrow\infty$.  Moreover, since
this is a lossless Hermitian system, all the resonant frequencies
are real.  But when the system is separated into a single Lorentz
oscillator coupled to a bath of harmonic oscillators, one can
discern energy flow from the Lorentz oscillator to the bath as shall
be shown by the following analysis.  One can show that the natural
mode of the ``lossy" Lorentz (DLS) oscillator becomes complex implying that the
natural mode decays with time.  To find
the natural modes of the coupled harmonic oscillators, the driving
term or the last term in the Hamiltonian in \eqref{qd:eq4} can be
ignored.


To this end, one can then transform the above into normal variables \cite{Tannoudji} by
letting
\begin{align}\label{qd:eq5}
\hat\zeta=\frac1{\sqrt{2}}\left(\hat a+\hat a^\dag\right),\quad
\hat\pi =\frac1{i\sqrt{2}}\left(\hat a-\hat a^\dag\right),\nonumber\\
\hat\zeta_j=\frac1{\sqrt{2}}\left(\hat b_j+\hat
b_j^\dag\right),\quad \hat\pi_j =\frac1{i\sqrt{2}}\left(\hat
b_j-\hat b_j^\dag\right)
\end{align}
where $\hat a$ and $\hat b_j$ represent the modes of the single
Lorentz oscillator and the $j$-th bath oscillator, respectively.
 Consequently, the Hamiltonian becomes
\begin{align}\label{qd:eq6}
{\hat H}_{PB} = &\frac12\hbar\omega_0 \left( \hat{a}\hat a^\dag +
\hat{a}^\dag\hat a\right) + \sum_j \frac12 \hbar\omega_j\left( \hat
b_j \hat b_j^\dag + \hat b_j^\dag \hat b_j\right)\nonumber\\
&+\sum_j\hbar\frac{\alpha_j^\zeta+\alpha_j^\pi}{\sqrt{2}}\left( \hat a \hat
b_j^\dag+\hat a^\dag \hat b_j\right)\nonumber\\
&+\sum_j\hbar\frac{\alpha_j^\zeta-\alpha_j^\pi}{\sqrt{2}}\left( \hat a \hat
b_j+\hat a^\dag \hat b_j^\dag\right)
\end{align}
If $\alpha^\zeta_j=\alpha^\pi_j$, then the last term above vanishes;
or one can make the rotating wave approximation that the last term
is rapidly varying, and hence, its contribution to the total
Hamiltonian is small.  Therefore, the final Hamiltonian becomes
\begin{align}\label{qd:eq7}
{\hat H}_{PB}= &\frac12\hbar\omega_0\left(\hat a \hat a^\dag+\hat
a^\dag\hat a \right) + \sum_j\frac12\hbar \omega_j\left(\hat b_j\hat
b_j^\dag+\hat b_j^\dag\hat b_j\right)\nonumber\\
&+\sum_j \tilde\gamma_j\hbar\left(\hat a \hat b_j^\dag + \hat a^\dag \hat
b_j\right)
\end{align}
where $\tilde\gamma_j=
\frac{\alpha_j^\zeta+\alpha_j^\pi}{\sqrt{2}}$.
The \EsOM\ for $\hat a$ and $\hat b_j$ can be easily derived using
the Heisenberg \EOM\, or
\begin{align}
\dot{\hat a}=\frac1{i\hbar}\left[\hat a,\hat H\right],\qquad
\dot{\hat b_j}=\frac1{i\hbar}\left[\hat b_j,\hat H\right]
\end{align}
They become \cite{Footnote2}
\begin{align}\label{qd:eq8}
\dot{\hat a}&=-i\left(\omega_0\hat a +\sum_j\tilde\gamma_j\hat b_j\right)\\
\dot{\hat b}_j &=-i\left(\omega_j\hat  b_j +\tilde\gamma_j \hat
a\right)\label{qd:eq9}
\end{align}
The above is a Hermitian system with no loss.  The corresponding
eigenmodes of the system correspond to lossless time-harmonic
oscillators of a Hermitian system.  If there are $N$
oscillators coupled together, there would be $N$ degrees of freedom
and this system of equations yields $N$ modes.  These equations
account for the coupling of the single Lorentz oscillator to the noise
bath oscillators, but not the coupling between the oscillators
within the noise bath.

However, if the initial condition is such that the starting states
of the harmonic oscillators in the bath are completely random,
%
%
it has been shown in \cite[eq. (6.89)]{HAUS}, \cite{TAN} and \cite{Lax} that
the leakage of energy from the single oscillator to the bath gives
rise to the decay of the amplitude of the harmonic oscillator.

\begin{figure}[h]
\includegraphics[width = 0.45\textwidth]{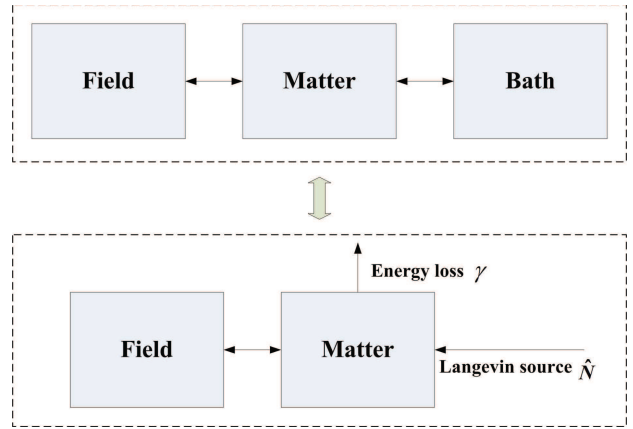}
\caption{\label{fig:2} The field-matter-bath coupled system. The effect of the bath is the introduction of loss and Langevin sources into the matter system, which further influences the field.}
\end{figure}

Following \cite{TAN} by defining $\hat b_j=\hat{\tilde b}_j
e^{-i\omega_j t}$, the second equation can be simplified as
\begin{align}\label{qd:eq13}%
\dot{\hat{ \tilde b}}_j =  -i \tilde\gamma_j\hat{  a} e^{i\omega_j
t}
\end{align}
The above equation can be integrated to yield
\begin{align}\label{qd:eq14}
\hat{\tilde b}_j(t) = -i\int_0^t \tilde\gamma_j\hat{ a}(\tau)
e^{i\omega_j\tau} d\tau +\hat{ \tilde b}_j(0)
\end{align}
Upon substituting the above into equation \eqref{qd:eq8}, and
exchanging the order of integration and summation, one arrives at
\begin{align}\label{qd:eq15}
\dot{\hat{ a}} = &-i\omega_0\hat{ a} - \int_0^t \sum_j
\tilde\gamma_j^2 \hat{a}(\tau) e^{-i\omega_j (t-\tau)} d\tau\nonumber\\
&-i\sum_j \tilde\gamma_j\hat{ \tilde b}_j(0) e^{-i\omega_j t}
\end{align}
The above can be thought of as a model for detailed balance, but it
has no closed-form expression for the terms.
%
%
So to obtain approximate analytic expressions for the terms, one can
study the summation terms inside the second term on the \RHS\ in
greater detail.  The summation above is given by
\begin{align}\label{qd:eq16a}
\sum_j \tilde\gamma_j^2 e^{-i\omega_j(t-\tau)}
\end{align}
It clearly peaks when $t=\tau$.  Moreover, if one assumes that
$\gamma_j$ is weakly dependent on $j$ so that it can be approximated
by $\tilde\gamma_j^2\approx\frac\eta{2\pi} \Delta \omega$ then
\begin{align}\label{qd:eq16}
\sum_j \tilde\gamma_j^2 e^{-i\omega_j(t-\tau)}\approx
\frac{\eta}{2\pi}\int_{-\infty}^\infty e^{-i\omega(t-\tau)}d\omega =
\eta \delta(t-\tau)
\end{align}
 The above equation now becomes
\begin{align}\label{qd:eq17}
\dot{\hat{ a}} = -i\omega_0\hat{ a }- \eta\hat{ a}
 - i\sum_j \tilde\gamma_j\hat{ \tilde
b}_j(0) e^{-i\omega_j t}
\end{align}
which is the same as that derived by the Laplace transform method in
the Appendix as $\hat{\tilde b}_j(0)=\hat{b}_j(0)$.  It is
interesting to note that the coupling to the bath introduces a
dissipation term given by $- \eta\hat{ a}$, but it is also augmented
by a source term given by the last term above.  The augmented source
term can be regarded as the Langevin source. In Figure \ref{fig:2},
the coupled system is sketched.

\section{Physical Interpretation}

One can explain the physical meaning of the above further.
Equations \eqref{qd:eq8} and \eqref{qd:eq9}, represent a Hermitian
lossless system of coupled oscillating modes.  All the resonant
modes of the Hermitian system can be proved to be real. However, the
second term on the \RHS\ of \eqref{qd:eq8} implies that $\hat a$ is
being driven by $\hat b_j$.  But $\hat b_j$ is being also bring
driven by $\hat a$: this point is being expressed by
\eqref{qd:eq14}.  The first term on the \RHS\ represents the back
action by $\hat a$ on $\hat b_j$, while the second term implies that
$\hat b_j$ will remain unchanged if $\hat a$ and $\hat b_j$ are not
coupled together.

In \eqref{qd:eq15}, the second term on the \RHS\ can be written as a
convolutional integral:
\begin{align}\label{pi:eq1}%
\hat a(t)\circledast B(t)
\end{align}
where $B(t)=\sum_j\tilde \gamma_j^2 e^{-i\omega_j t}$.  It is seen
that $B(t)$ is the sum of many oscillators with different
frequencies. This term eventually leads to the loss term in
\eqref{qd:eq17}, but the loss is caused by the destructive
interference or non-time-reversibility of the oscillators in the
bath. That it becomes a loss term that ``siphons" energy from the
oscillator to the bath is only valid in the ensemble average sense.
The last term in \eqref{qd:eq17}, on the other hand, ``pumps" energy
back into the $\hat a$ oscillator. But it is seen that the last term
is again a sum of incoherent oscillators with initial values $\hat
b_j(0)$ which is random.

The initial value $\hat b_j(0)$ is very much related to the
temperature of the noise bath:  The higher the temperature, the
larger the initial values.  Also, this term is ``white" (as
in white noise) compared to the term that siphons energy off the
$\hat a$ oscillator.  By energy conservation, at thermal equilibrium, these two energies,
siphoned energy and pumped energy, should be equal to each other,
but they are equal only in the ensemble average sense.

Because of this physical interpretation, the last term in
\eqref{qd:eq17} can be expressed as the Langevin noise source,
namely,
\begin{align} \label{pi:eq2}
\dot{\hat{ a}} = -i\omega_0 \hat{a}- \eta\hat{ a} + \hat F(t)
\end{align}
where
\begin{align}\label{pi:eq4}%
\hat F(t)= - i\sum_j \tilde\gamma_j\hat{ b}_j(0) e^{-i\omega_j t}
\end{align}
By the same token, the creation operator equivalence of the above is
\begin{align}\label{pi:eq3} \dot{\hat{ a}}^\dag = i\omega_0 \hat{a}^\dag-
\eta\hat{ a}^\dag + \hat F^\dag(t)
\end{align}
The above system cannot be proven to be Hermitian, but it has
descended from a Hermitian system with asymptotic approximations. So
it should be quasi-Hermitian, or Hermitian in the average sense. Or
one can envision that the Langevin sources produce a response that
compensates the loss due to coupling of the system to the noise
bath.  This is similar in spirit to the fluctuation dissipation
theorem \cite{Johnson,Nyquist,CallenWelton} where the loss of
energy from the system to the noise bath in thermal equilibrium is
compensated by the back-coupling of energy back from the noise bath
to the system.

Furthermore,  it can be shown that the Langevin sources are highly
uncorrelated in time such that \cite{HAUS,TAN}
\begin{align}\label{pi:eq5}%
\langle [\hat F(t),\hat F^\dag (t')]\rangle &=\sum_j |\gamma_j|^2
e^{i\omega_j(t'-t)}\langle [\hat b_j(0),\hat b_j^\dag(0)]\rangle
\nonumber\\
&=2\eta \delta(t-t')
\end{align}
where the angular bracket implies ensemble average.   And more
importantly, the commutator between the $\hat a$ and $\hat a^\dag$
is preserved, as it was in the original Hermitian system. The real
and imaginary parts of operators $\hat a$ and $\hat a^\dag$ are
related to the quantum observables $\hat p$ and $\hat q$.  The
commutator of $\hat p$ and $\hat q$ is related to the commutator of
$\hat a$ and $\hat a^\dag$. As has been seen before, these
commutators are important for inducing the quantum Hamilton
equations of motion. The loss of these commutators would have meant
that the quantum \EsOM\ are not preserved.  It also means that
energy is not conserved. This would have been bizzare, as it means
that physical laws are not preserved.

One can define a Langevin noise operator such that
\begin{align}\label{8.6}
\hat F(t)=\sqrt{2\eta} \hat f(t)
\end{align}
then
\begin{align}\label{8.6b}
\left[\hat f(t),\hat f^\dag(t')\right]=\delta(t-t')
\end{align}
implying that the Langevin source is ``white".

Some further remarks are in order here. We have elected to follow the field-matter-bath model of Huttner and Barnett \cite{Huttner&Barnett}, in which a specific species of Lorentz oscillator field is chosen to represent the most prominent physical resonance of the material. Loss in the material to other degrees of freedom are then introduced through the bath oscillators, which correspond to other vibrations or excitations of the material not strongly coupled to the driving field. This description, as we have seen, holds in the classical as well as quantum mechanical cases. In the classical case it can model the loss term in Equation (\ref{hepv:eq1}), in the quantum case it leads to the quantum loss and Langevin sources.

There is an alternative viewpoint in which no explicit mention of a bath is invoked. The matter is treated in effect as a bath and the fields are coupled to not a single species, but many, many species of Lorentz oscillators everywhere in space. Such is the approach taken in the work of Philbin \cite{Philbin}, for example. We venture to say that the two viewpoints are the same. Whether one introduces a huge variety of Lorentz oscillators with differing frequencies at the outset, or later through coupling, it is a bath of many oscillators that produces the quantum loss and Langevin sources. In fact, using the technique of Fano diagonalization \cite{Huttner&Barnett, Fano} we can easily turn the coupled oscillators in Equation  (\ref{qd:eq7}) into independent oscillators of differing frequencies, all of which are coupled to the field explicitly.

The usefulness of the two viewpoints depends on the application. We wish to emphasize the quantum analogy of classical loss here, hence the field-matter-bath model. This model tailors toward a specific type of dispersion relation, as will be seen in the discussions to follow.

\section{Connecting Back with $\hat\zeta$, $\hat\pi$, and macroscopic quantities $\hat {\v
P}$ and $\hat{\vg\Pi}$}

In order to connect this quantum loss back with the macroscopic
Maxwell's equations, one needs to first connect back with the
$\zeta$ and $\pi$ variables model in \eqref{qd:eq5}. Therefore, one
arrives at the following equation pair:
\begin{align}\label{cb:eq1}
\dot{\hat\zeta}=\omega_0\hat \pi -\eta\hat \zeta +2\sqrt{\eta}\hat
f_R\\\label{cb:eq2} \dot{\hat\pi}=-\omega_0 \hat\zeta -\eta\hat \pi
+2\sqrt{\eta}\hat f_I
\end{align}
where from \eqref{8.6}, it follows that $\hat f(t)=\hat f_R(t)+i
\hat f_I(t)$ where $\hat f_R(t)$ and $\hat f_I(t)$ are Hermitian
operators, or respectively, the ``real" and ``imaginary" parts of
the $\hat f(t)$ operator. It can be shown that
\begin{align}\label{pi:eq3a}%
\langle [\hat f_R(t),\hat f_I(t')]\rangle= i\frac12 \delta(t-t')
\end{align}
%
The above is the result of the single Lorentz oscillator.  It needs to
be connected to the macroscopic Lorentz oscillator in a material
medium which is due to a cluster of Lorentz oscillators. Also, the
above is written in terms of dimensionless coordinate $\zeta$ and
$\pi$.
%
Here, the polarization density is normalized, and the collection of
Lorentz oscillators is still a simple harmonic oscillators.  Hence,
the macroscopic harmonic oscillator can be connected to the single
harmonic oscillator, as in \eqref{qd:eq2} by equating
\begin{align}
\hbar\omega_0\hat \zeta^2 = f\hat{\v P}^2
\end{align}
Initially, $\hat\zeta$ was a dimensionless quantity.  By forcing
this equality, $\hat\zeta$ now is a dimensioned quantity to be
commensurate with the above equality.  Hence, the macroscopic
quantity, $\hat{\v P}$ can be related to the microscopic quantity
$\hat\zeta$ by a multiplicative constant, so are the other
microscopic quantities such as $\pi$ and the Langevin source terms
$\hat f_R$ and $\hat f_I$. From the above, one connects $\hat
\zeta$, the $i$-th component of the polarization density of a medium
is given by
\begin{align}\label{cb:eq7}%
\hat P_i=q\sqrt{\frac{n\hbar}{m\omega_0\epsilon}}\hat \zeta_i
\end{align}
where the subscript $i$ here implies $x,y,z$ components, $n$ is the
electron density per unit volume, and $q$ is assumed positive.  The
polarization density is normalized such that $P_i^2$ is energy
density. Similarly,
\begin{align}\label{cb:eq7aa}%
\hat \Pi_i=q\sqrt{\frac{n\hbar}{m\omega_0\epsilon}}\hat \pi_i
\end{align}
Multiplying \eqref{cb:eq1} and \eqref{cb:eq2} by the constant
$q\sqrt{\frac{n\hbar}{m\omega_0\epsilon}}$ yields
\begin{align}
\dot{\hat P}_i=\omega_0 \hat \Pi_i-\eta \hat P_i +\hat F_{R,i}\\
\dot{\hat \Pi}_i=-\omega_0 \hat P_i-\eta \hat \Pi_i +\hat F_{I,i}
\end{align}
%
where
\begin{align}\label{cb:eq9}
\hat F_{X,i}=q
\sqrt{\frac{n\hbar}{m\omega_0\epsilon}}2\sqrt{\eta}\left({\hat
f}_{X,i}\right)
\end{align}
where $X$ is either $R$ or $I$.  Putting the $x$, $y$, $z$
components of the above together to make a vector field, the above
equations become
\begin{align}\label{cb:eq10a}
\dot{\hat {\v P}}=\omega_0 \hat {\vg\Pi}-\eta \hat {\v P} +\hat {\v
F}_{R}\\\label{cb:eq10b} \dot{\hat {\vg\Pi}}=-\omega_0 \hat {\v
P}-\eta \hat {\vg\Pi} +\hat {\v F}_{I}
\end{align}
The above represents the damped oscillating mode of the lossy Lorentz
oscillator when it is coupled to noise bath.  The effect of the
noise bath is to induce dissipation as well as introducing Langevin
sources to compensate for the loss.  This makes the system
quasi-Hermitian in the ensemble average sense.  A new correlation
between the Langevin source terms can be easily established as
follows:
\begin{align}\label{cb:11c}
\left\langle\left[\hat{\v F}_R,\hat{\v F}_I\right]\right\rangle
=i2\eta\frac{\omega_p^2\hbar }{\omega_0}\delta(t-t')\hat{\dyad I}
\end{align}

When the driving field $\v E$ is turned back on, the second equation
above has to be modified accordingly to yield
\begin{align}\label{cb:eq11}
\dot{\hat {\vg\Pi}}=-\omega_0 \hat {\v P}-\eta \hat {\vg\Pi} +\hat
{\v F}_{I}+\frac{\omega_p^2}{\omega_0}\hat{\v E}
\end{align}
The above equation, together with \eqref{cb:eq10a}, and the quantum
Maxwell's equations, can be solved in tandem to yield the \EsOM\ for
a dissipative quantum electromagnetic system.

The above quantum operator equations for the lossy Lorentz
oscillator can be solved in tandem with the rest of the quantum
electromagnetic equations.
\begin{align}\label{cb:eq6a}
 \dot {\hat{\v H}}\prt  &= -\Curl\hat{\v
E}\prt\\\label{cb:eq7a}%
 \dot{ \hat{\v E}}\prt  &=
\Curl\hat{\v H}\prt -\dot{\hat{\v P}}\prt
\end{align}
They constitute the quantum electromagnetic equations for a lossy
system coupled to a noise bath.  The coupling to the noise bath
gives rise to Langevin sources that are needed to retain their
Hermitian or lossless nature in the average sense.

Even though the above equations have been derived by coupling the
Lorentz oscillator to a noise bath of harmonic oscillators, they
could also have been derived axiomatically.  For instance, one can
postulate the noise bath to have the statistical property of a white
noise as indicated by \eqref{8.6b}.  Then as shown in the Appendix,
the commutator is preserved in the ensemble average sense.

\section{Connection to FDT and the Work of Welsch's Group}

In this section, the connection of the Langevin sources derived here
will be connected to the Langevin sources in Welsch's group.  In his
group, the commutator for the Langevin noise current has been
motivated by the fluctuation dissipation theorem (FDT).  The
connection of this work to FDT and hence, the work of Welsch's group
will be shown.  First, the low-loss case will be considered
followed by the high-loss case.

\subsection{Low Loss Case}

From the above equations \eqref{cb:eq10a} to \eqref{cb:eq11}, one
can derive that
\begin{align}\label{cb:eq5c}%
\ddot {\hat{\v P}}\prt&=-\omega_0^2\hat{ \v P}\prt
-\eta\omega_0\hat{\vg\Pi}\prt-\eta\dot{\hat{\v
P}}\prt\notag\\&\qquad +\omega_0\hat{\v F}_I\prt+\dot{\hat{\v
F}}_R\prt+\omega_p^2\hat{\v E}\prt
\end{align}
In the above, $\omega_0\hat{\vg\Pi}\approx\dot{\hat{\v P}}$ when the
loss is low. Then it can be reduced to
\begin{align}\label{cb:eq5b}%
\ddot {\hat{\v P}}\prt+\gamma \dot{\hat{\v P}}\prt+\omega_0^2\hat{
\v P}\prt \doteq \omega_p^2 \hat{\v E}\prt + \hat{\v N}\prt
\end{align}
where $\gamma=2\eta$, and $\hat{\v N}=\omega_0\hat{\v
F}_I+\dot{\hat{\v F}}_R$ is the Langevin source.  The $i$-th
component of the Langevin source only contributes to the $i$-th
component of the above equation.

The above illustrates the interesting notion that the lossy Lorentz
oscillator is being driven by the field $\v E$ as well as the
Langevin source $\v N$.  It also illustrates the notion that loss in
Maxwell's equations come from coupling to a noise bath that is
formed by other harmonic oscillators.  When there is no material,
there is no loss. \textit{That explains why a photon due to the free field
in vacuum can travel through the vast galaxy without being absorbed}.



From \eqref{cb:eq6a}, \eqref{cb:eq7a}, and \eqref{cb:eq5b} and with
Fourier transform, one gets the quantized vector wave equation of
electric field in the frequency domain, i.e.
\begin{equation}\label{eq_fdt1}
\nabla\times\nabla\times\hat{\mathbf{E}}(\mathbf{r},\omega)-\omega^2\epsilon(\mathbf{r},\omega)\hat{\mathbf{E}}(\mathbf{r},\omega)=i\omega
\hat{\mathbf{j}}_n(\mathbf{r},\omega)
\end{equation}
where
\begin{align}\label{eq_fdt1_add}
\epsilon(\mathbf{r},\omega)=1+\frac{\omega_p^2}{\omega_0^2-\omega^2-i\omega
\gamma}
\end{align}
is the permittivity for the lossy and dispersive media satisfying the classical DLS model.
Here, $\omega_p$, $\omega_0$, and $\gamma$ can be functions of $\v
r$.  Moreover, the noise current $\hat{\mathbf{j}}_n$ can be
expressed as
\begin{equation}\label{eq_fdt2}
\hat{\mathbf{j}}_n(\mathbf{r},\omega)=-i\omega\frac{\hat{\mathbf{N}}(\mathbf{r},\omega)}{\omega_0^2-\omega^2-i\omega
\gamma}
\end{equation}

From \eqref{cb:11c}, the correlation between the Langevin source terms in frequency domain can be written as
\begin{align}\label{eq_fdt2_add}
\langle[\hat{\mathbf{F}}_R(\omega), \hat{\mathbf{F}}_I^{\dag}(\omega')]\rangle&=\frac{1}{(2\pi)^2}\iint\langle[\hat{\mathbf{F}}_R(t), \hat{\mathbf{F}}_I(t')]\rangle e^{i\omega t-i\omega't'}dtdt'\notag\\
&=\frac{1}{(2\pi)^2}\int i \gamma \frac{\omega_p^2\hbar}{\omega_0}e^{i(\omega-\omega')t} dt\notag\\
&=\frac{i\gamma}{2\pi}\frac{\omega_p^2\hbar}{\omega_0}\delta(\omega-\omega')
\end{align}

\noindent Then, the commutation relation of the Langevin noise source
$\hat{\mathbf{N}}$ can be obtained
\eqref{pi:eq3a}
\begin{equation}\label{eq_fdt3}
\langle[\hat{\mathbf{N}}(\mathbf{r},\omega),\hat{\mathbf{N}}^{\dag}(\mathbf{r},\omega')]\rangle
=\frac{\omega_p^2\hbar\gamma(\mathbf{r})\omega}{\pi}\delta(\omega-\omega')
\end{equation}
\noindent By using \eqref{eq_fdt2} and \eqref{eq_fdt3}, the
commutation relation of the noise current is of the form
\begin{align}\label{eq_fdt4}
\langle[\hat{\mathbf{j}}_n(\mathbf{r},\omega),\hat{\mathbf{j}}_n^{\dag}(\mathbf{r},\omega')]\rangle
&=
\omega^2\frac{[\hat{\mathbf{N}}(\mathbf{r},\omega),\hat{\mathbf{N}}^{\dag}(\mathbf{r},\omega')]}{(\omega_0^2-\omega^2)^2+\omega^2\gamma(\mathbf{r})^2}\nonumber\\
&=\frac{\hbar\omega^2}{\pi}
\frac{\omega_p^2\omega\gamma(\mathbf{r})}{(\omega_0^2-\omega^2)^2+\omega^2\gamma(\mathbf{r})^2}\delta(\omega-\omega')
\end{align}
\noindent The conductivity of the lossy and dispersive media is
given by
\begin{equation}\label{eq_fdt5}
\sigma(\mathbf{r},\omega)=\omega \Im
m\left\{{\epsilon(\mathbf{r},\omega)}\right\}=
\omega\frac{\omega_p^2\omega\gamma(\mathbf{r})}{(\omega_0^2-\omega^2)^2+\omega^2\gamma(\mathbf{r})^2}
\end{equation}
Hence, the commutation relation Eq. \eqref{eq_fdt4} can be
simplified as
\begin{equation}\label{eq_fdt6}
\langle[\hat{\mathbf{j}}_n(\mathbf{r},\omega),\hat{\mathbf{j}}_n^{\dag}(\mathbf{r},\omega')]\rangle=\frac{\hbar\omega}{\pi}\sigma(\mathbf{r},\omega)\delta(\omega-\omega')
\end{equation}
The above agrees with Eq. (18) in the paper from Welsch's group
\cite{DungEtal}, Eq. (3.54) in Scheel and Buhmann
\cite{Scheel&Buhmann}, and the fluctuation dissipation theorem.

\def\VWF{vector wave function}
\def\PBC{periodic boundary condition}

\subsection{High Loss Case}

From \eqref{cb:eq10a} and \eqref{cb:eq11}, the polarization density
$\mathbf{P}$ in the frequency domain is given by
\begin{align}\label{hl:eq1}
\mathbf{\hat{P}}(\mathbf{r},\omega)=\frac{\omega_p^2\mathbf{\hat{E}(\mathbf{r},\omega)}+\omega_0\mathbf{\hat{F}}_I(\mathbf{r})+
\left[\eta(\mathbf{r})-i\omega\right]\mathbf{\hat{F}}_R(\mathbf{r})}{\left[\eta(\mathbf{r})-i\omega\right]^2+\omega_0^2}
\end{align}
where it is reminded that $\eta=\gamma/2$.   Then, following the same
procedure, one gets the same quantized vector wave equation
\eqref{eq_fdt1} with the permittivity
\begin{align}\label{hl:eq2}
\epsilon(\mathbf{r},\omega)&=1+\frac{\omega_p^2}{\left[\eta(\mathbf{r})-i\omega\right]^2+\omega_0^2}\notag\\
&=1+\frac{\omega_p^2}{\omega_0^2-\omega^2-i\omega \gamma+\gamma^2/4}
\end{align}
and the noise current
\begin{align}\label{hl:eq3}
\mathbf{\hat{j}}_n(\mathbf{r},\omega)=\frac{-i\omega\omega_0\mathbf{\hat{F}}_I(\mathbf{r})-i\omega
\left[\eta(\mathbf{r})-i\omega\right]\mathbf{\hat{F}}_R(\mathbf{r})}{\left[\eta(\mathbf{r})-i\omega\right]^2+\omega_0^2}
\end{align}
Using \eqref{eq_fdt2_add}, the commutation relation of the noise current
can be written as
\begin{align}\label{hl:eq4}
&\left\langle \left[\mathbf{\hat{j}}_n(\mathbf{r},\omega),\mathbf{\hat{j}}_n^{\dag}(\mathbf{r},\omega')\right]\right\rangle\notag\\
&=\frac{2\hbar\eta\omega^3\omega_p^2}{\pi}
\frac{1}{(\omega_0^2+\eta^2-\omega^2)^2+4\eta^2\omega^2}
\delta(\omega-\omega')
\end{align}
According to \eqref{hl:eq2}, the imaginary part of permittivity is
of form
\begin{align}\label{hl:eq5}
\Im
m\{\epsilon(\mathbf{r},\omega)\}=\frac{2\eta\omega\omega_p^2}{(\omega_0^2+\eta^2-\omega^2)^2+4\eta^2\omega^2}
\end{align}
Finally, the commutation relation is simplified to
\begin{align}\label{hl:eq6}
\left\langle\left[\mathbf{\hat{j}}_n(\mathbf{r},\omega),\mathbf{\hat{j}}_n^{\dag}(\mathbf{r},\omega')\right]\right\rangle
&= \frac{\hbar\omega}{\pi}\omega\Im
m\{\epsilon(\mathbf{r},\omega)\}\delta(\omega-\omega')\notag\\
&=\frac{\hbar\omega}{\pi}\sigma(\mathbf{r},\omega)\delta(\omega-\omega')
\end{align}
Interestingly, the commutation relation still maintains the same
form as the low loss case \eqref{eq_fdt6}, although the classical
DLS model breaks down as shown in \eqref{hl:eq2}. Comparing the
quantum equations of motion \eqref{cb:eq10a} and \eqref{cb:eq11} to the
corresponding classical counterparts \eqref{hepv:eq4} and
\eqref{hepv:eq5}, the Langevin noise is introduced to the two
complementary quantum variables simultaneously.

\section{Application to two-level quantum system}
We consider a two-level quantum system interacting with fluctuating electromagnetic fields in arbitrary lossy and dispersive media. The wave function in the interaction (Dirac) picture satisfies the following equation
\begin{align}\label{qs:eq2}
i\hbar\frac{\partial \Psi_I(t)}{\partial t}=\hat{\mathbf{V}}_{I}(t)\Psi_I(t)
\end{align}
where $\hat{\mathbf{V}}_{I}(t)$ is the interaction Hamiltonian operator. Using dipole approximation, we have
\begin{align}\label{qs:eq3}
\hat{\mathbf{V}}_{I}(t)=-\mathbf{\hat{\mu}}_{I}(t)\mathbf{\hat{E}}_{I}(t)
\end{align}
where $\mathbf{\hat{\mu}}_{I}$ and $\mathbf{\hat{E}}_{I}$ are, respectively, the dipole moment and electric field operators in the interaction picture. $\mathbf{\hat{E}}_{I}(t)$ is a short notation of $\mathbf{\hat{E}}_{I}(\mathbf{r}_0,t)$, where $\mathbf{r}_0$ is the spatial location of the artificial atom modeled as the two-level quantum system.

The solution to \eqref{qs:eq2} is of form
\begin{align}\label{qs:eq4}
\Psi_I(t)=\exp\left(\frac{1}{i\hbar}\int_0^t \hat{\mathbf{V}}_{I}(t)dt\right)|\Psi_S(0)\rangle
\end{align}
where $|\Psi_S(0)\rangle$ is the initial wave function in the Schr\"odinger picture. Also, it is easy to show
\begin{align}\label{qs:eq5}
\mathbf{\hat{\mu}}_{I}(t)&=\exp(i\mathbf{\hat{H}}_{0}t/\hbar)\mathbf{\hat{\mu}}_{S}\exp(-i\mathbf{\hat{H}}_{0}t/\hbar)\notag\\
&=\exp(i\omega_{eg}t)\mathbf{\mu}_{eg}|e\rangle\langle g|+\exp(-i\omega_{eg}t)\mathbf{\mu}_{ge}|g\rangle\langle e|\notag\\
&=\exp(i\omega_{eg}t)\mathbf{\hat{\mu}}_{eg}+\exp(-i\omega_{eg}t)\mathbf{\hat{\mu}}_{ge}
\end{align}
where $\mathbf{\hat{H}}_{0}$ is the superposition of the atomic Hamiltonian and the field-matter-bath Hamiltonian. Moreover, $|g\rangle$ and $|e\rangle$ represent the ground and excited states of the two-level system, respectively. $\omega_{eg}=\omega_e-\omega_g$ and $\mathbf{\mu}_{eg}$ are the corresponding transition frequency and dipole moment, respectively.

Substituting \eqref{qs:eq5} and \eqref{qs:eq3} into \eqref{qs:eq4}, we arrive at
\begin{align}\label{qs:eq6}
\Psi_I(t)&=\exp\left(\frac{\mathbf{\hat{\mu}}_{ge}}{i\hbar}\int_{-\infty}^{\infty}\int_0^t \exp(-i(\omega_{eg}+\omega)t)\mathbf{\hat{E}}_{I}(\omega)dtd\omega\right.\notag\\
&+\left.\frac{\mathbf{\hat{\mu}}_{eg}}{i\hbar}\int_{-\infty}^{\infty}\int_0^t \exp(i(\omega_{eg}-\omega)t)\mathbf{\hat{E}}_{I}(\omega)dtd\omega\right)|\Psi_S(0)\rangle\notag\\
&=\exp\left(\frac{\mathbf{\hat{\mu}}_{ge}}{i\hbar}\int_{-\infty}^{\infty}\frac{1-\exp(-i(\omega_{eg}+\omega)t)}{i(\omega_{eg}+\omega)}\mathbf{\hat{E}}_{I}(\omega)d\omega\right.\notag\\
&+\left.\frac{\mathbf{\hat{\mu}}_{eg}}{i\hbar}\int_{-\infty}^{\infty}\frac{\exp(i(\omega_{eg}-\omega)t)-1}{i(\omega_{eg}-\omega)}\mathbf{\hat{E}}_{I}(\omega)d\omega\right)|\Psi_S(0)\rangle
\end{align}

With the help of the wave function \eqref{qs:eq6}, the expectation value of an arbitrary operator can be obtained. For example, the average population of the excited state is given by
\begin{align}\label{qs:eq7}
\left\langle|e\rangle\langle e|\right\rangle=\left\langle\hat{\sigma}_e\right\rangle=\langle\Psi_I(t)|\hat{\sigma}_e|\Psi_I(t)\rangle
\end{align}
We will study how to rigorously calculate \eqref{qs:eq7} by a fast and efficient numerical algorithm in near future.

In the following, spontaneous emission of the two-level system will be solved by a linear approximation of \eqref{qs:eq6}. The initial wave function ($t=0$) in the Schr\"odinger picture is assumed to
\begin{align}\label{qs:eq1}
|\Psi_S(0)\rangle=|g\rangle |0\rangle
\end{align}
where $|0\rangle$ means the zero-photon state. From \eqref{qs:eq6} and \eqref{qs:eq7}, the wave function and the average population can be approximated respectively as
\begin{align}\label{qs:eq8}
\Psi_I(t)\approx&|g\rangle|0\rangle+\frac{\mathbf{\mu}_{eg}|e\rangle}{i\hbar}\int_{-\infty}^{\infty}\mathbf{\hat{E}}_{I}(\omega)|0\rangle\notag\\
&\frac{\exp(i(\omega_{eg}-\omega)t)-1}{i(\omega_{eg}-\omega)} d\omega
\end{align}
and
\begin{align}\label{qs:eq9}
\left\langle\hat{\sigma}_e\right\rangle\approx&\frac{|\mathbf{\mu}_{eg}|^2}{\hbar^2}\int_{-\infty}^{\infty}\int_{-\infty}^{\infty} \frac{\exp(-i(\omega_{eg}-\omega')t)-1}{-i(\omega_{eg}-\omega')} \notag\\
&\frac{\exp(i(\omega_{eg}-\omega)t)-1}{i(\omega_{eg}-\omega)}\langle0|\mathbf{\hat{E}}^{\dag}_{I}(\omega')\mathbf{\hat{E}}_{I}(\omega)|0\rangle d\omega' d\omega \notag\\
=&\frac{|\mathbf{\mu}_{eg}|^2}{\hbar^2}\int_{-\infty}^{\infty}\left|\frac{\exp(i(\omega_{eg}-\omega)t)-1}{i(\omega_{eg}-\omega)}\right|^2\notag\\ &\langle0|\mathbf{\hat{E}}^{\dag}_{I}(\omega)\mathbf{\hat{E}}_{I}(\omega)|0\rangle d\omega\notag\\
=&\frac{|\mathbf{\mu}_{eg}|^2}{\hbar^2}\int_{-\infty}^{\infty}\left|\frac{\exp(i(\omega_{eg}-\omega)t)-1}{i(\omega_{eg}-\omega)}\right|^2\notag\\ &\langle0|\mathbf{\hat{E}}^{\dag}_{S}(\omega)\mathbf{\hat{E}}_{S}(\omega)|0\rangle d\omega
\end{align}
where the field operators are uncorrelated at different frequencies.

If the interaction time between the field-matter-bath system and atomic system is long enough, then we have
\begin{align}\label{qs:eq10}
\left|\frac{\exp(i(\omega_{eg}-\omega)t)-1}{i(\omega_{eg}-\omega)}\right|^2_{t\rightarrow\infty}=2\pi t\delta(\omega-\omega_{eg})
\end{align}
Finally, the average population of the excited state is approximately simplified to
\begin{align}\label{qs:eq11}
\left\langle\hat{\sigma}_e\right\rangle\approx \frac{2\pi t|\mathbf{\mu}_{eg}|^2}{\hbar^2}\langle0|\mathbf{\hat{E}}^{\dag}_{S}(\omega_{eg})\mathbf{\hat{E}}_{S}(\omega_{eg})|0\rangle
\end{align}
The spontaneous emission rate is given by
\begin{align}\label{qs:eq12}
\gamma&=\frac{d\left\langle\hat{\sigma}_e\right\rangle}{dt}=\frac{2\pi |\mathbf{\mu}_{eg}|^2}{\hbar^2}\langle0|\mathbf{\hat{E}}^{\dag}_{S}(\omega_{eg})\mathbf{\hat{E}}_{S}(\omega_{eg})|0\rangle\notag\\
&=\frac{2\pi |\mathbf{\mu}_{eg}|^2}{\hbar^2}\langle0|\mathbf{\hat{E}}_{S}^{+}(\omega_{eg})\mathbf{\hat{E}}_{S}^{-}(\omega_{eg})|0\rangle
\end{align}
where $\mathbf{\hat{E}}_{S}^{+}$ and $\mathbf{\hat{E}}_{S}^{-}$ are the E-field operators with the positive and negative frequencies, respectively. The antinormal ordering operator $\mathbf{\hat{E}}_{S}^{+}\mathbf{\hat{E}}_{S}^{-}$ represents the probability for photon
emission \cite{Novotny&Hecht}.

According to fluctuation-dissipation theorem and quantum statistics \cite{Scheel&Buhmann}, the (thermal) expectation value of the electromagnetic fields is of form
\begin{align}\label{qs:eq13}
\langle0|\mathbf{\hat{E}}_{S}^{+}(\omega_{eg})&\mathbf{\hat{E}}_{S}^{-}(\omega_{eg})|0\rangle=
\langle0|\mathbf{\hat{E}}_{S}^{+}(\mathbf{r}_0,\omega_{eg})\mathbf{\hat{E}}_{S}^{-}(\mathbf{r}_0,\omega_{eg})|0\rangle \notag\\
&=\frac{\hbar\omega_{eg}^2}{\pi c^2\epsilon_0}[\bar{n}(\omega_{eg},T)+1]
\Im \overline{\mathbf{G}}(\mathbf{r}_0,\mathbf{r}_0,\omega_{eg})
\end{align}
where $\bar{n}(\omega_{eg},T)$ is the average thermal photon number. Particularly, the dyadic Green's function $\overline{\mathbf{G}}$ in the inhomogeneous, dispersive, and lossy media can be numerically calculated by classical computational electromagnetics methods \cite{Pengfei, Yongpin}.

\begin{figure}[h]
\includegraphics[width = 0.45\textwidth]{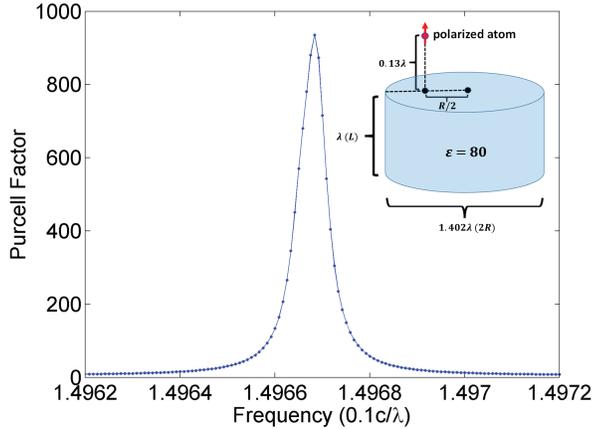}
\caption{\label{fig:3} Spontaneous emission enhancement of a polarized atom/molecule placed above a dielectric cylinder that supports a supercavity mode.}
\end{figure}

Here, we calculate the spontaneous emission from a polarized atom/molecule placed above a dielectric cylinder with the radius of 0.701 wavelength, height of 1 wavelength, and relative permittivity of 80. The cylinder supports a supercavity mode or bound states in the continuum, which is the interference of the Mie resonance (along the azimuthal direction) and Fabry-Perot resonance (along the vertical direction) \cite{Kivshar}. The atom is placed asymmetrically with respect to the cylinder axis for avoiding the nodal line of the supercavity mode. The spontaneous emission of the atom can be greatly enhanced with a Purcell factor close to $1000$, as shown in Fig. \ref{fig:3}.

\section{Conclusion}

We have presented a model for lossy, dispersive electromagnetic
medium that is valid in the quantum regime.  The medium is
dispersive because of the coupling of the Maxwellian free fields to
a set of Lorentz oscillators.  This can be regarded as field-matter
coupling in the parlance of previous work \cite{Huttner&Barnett}.
Furthermore, loss is induced in the quantum system by coupling to a
collection of simple harmonic oscillators to model noise in a
phenomenological manner.

The dispersion comes about because these Lorentz oscillators are
sluggish, and their dipole moments cannot be turned on (or off)
instantaneously.  The coupled system of the Lorentz oscillators to
the Maxwellian free field is quantized rigorously here,
corresponding to the quantization of a dispersive electromagnetic
system.  Such quantization is achieved without the need for mode
decomposition or diagonalization of the system. Also, such
quantization of the coupled system between free fields and matter has
not been seen before using this approach.
This is advantageous in some systems where
finding the normal modes could be a numerically intensive endeavor.

In this work, the coupling of field to matter consists of only one
species of Lorentz oscillators. The generalization to the
multi-species oscillators case is straightforward and will be shown
in our future work.

Also, the loss of the quantum system is obtained by coupling the single
oscillator to a bath; the bath induces loss in the single Lorentz
oscillator, making its resonance frequency complex.  The effect of
the noise bath is to shift the resonant frequency of the Lorentz
oscillator from being real to a complex number. Meanwhile, the
presence of loss requires the appearance of Langevin sources causing
the whole system to be energy conserving or quasi-Hermitian in the
ensemble average sense.
Hence, the loss model is considerably simpler allowing for analytic
solution to elucidate the physics behind the loss mechanism.  One
can also observe the one-way flow of energy from the host quantum
system consisting of the Lorentz oscillator to the noise bath.
Moreover, the \EsOM\ are considerably simpler and closer to the
classical model. The proximity of the model to classical model
allows the ease to incorporate computational electromagnetics
methods \cite{FEACEM} to solve future quantum problems.
The appearance of Langevin sources is commensurate with the physics
of the fluctuation dissipation theorem
\cite{Johnson,Nyquist,CallenWelton, Milonni, Rytov}:  at thermal equilibrium with a
noise bath, energy is lost from the quantum system to the bath, but
energy is also returned to the quantum system from the noise bath.
A more complicated noise model as expounded in \cite{Kira&Koch} can be assumed.
There, more complicated physical processes such as coupling to phonons and
electron collisions can give rise to dissipation but this is beyond
the scope of this work.


In the previous work involving the coupling between the free-field,
matter, and noise bath, the mode decomposition of the entire coupled
system is achieved with Fano diagonalization, giving rise to modes
which are called polaritons \cite{Huttner&Barnett}.
But here, no diagonalization is necessary, and the resulting quantum
equations of motion resemble the classical equations of motion
involving Lorentz oscillators.  It is hoped that this will enable a
simpler model for dissipative quantum electromagnetic systems as
well as future quantum technologies.

The ability to model quantum dissipation is important to determine
the coherence time of a quantum system.  This is especially
important in the design of quantum computers where the coherence
between qubits (quantum bits) or artificial atoms has to be
maintained.
This model also points out that in a pure vacuum, there could be no
quantum dissipation unless material media are present.  As
aforementioned, this explains why photons can traverse gigantic
distances in our universe.

\section{Acknowledgement}
We are grateful to Dr. Liping YANG of Purdue U for pointing out Louisell's book to us.

\appendix
\section{Energy Stored in the Field}\label{app:1}

Neglecting the $1/2$ factor, the Hamiltonian for the vector
potential is
\begin{align}
\mathscr{H}_{A,0} =\v \Pi_A^2 +\left(\Curl \v A\right)^2 +
\left(\Div \v A\right)^2=\v \Pi_A^2 +\v B^2 + \Pi_\Phi^2
\end{align}
where $\v \Pi_A=\dot {\v A}$, $\Div\v A=-\dot {\Phi}$, and $\v
B=\Curl \v A$.
\begin{align}
\mathscr{H}_{\Phi,0} = \Pi_\Phi^2 +\left(\nabla \Phi\right)^2
\end{align}
The subscript ``0" is used to indicate that these Hamiltonians are
the stand-alone Hamiltonians where coupling with the polarization
current is not accounted for. Therefore,
\begin{align}
\mathscr{H}_{A,0}-\mathscr{H}_{\Phi,0} =\v \Pi_A^2 +\v B^2 -
\left(\nabla \Phi\right)^2
\end{align}
But
\begin{align}
\mathscr{H}_{F,0} &= \v E^2+\v B^2 =\left(\v\Pi_A+\nabla\Phi\right)^2
+ \v B^2\nonumber\\
&=\v \Pi_A^2 + \left(\nabla \Phi\right)^2+2\nabla\Phi\cdot\v
\Pi_A+\v B^2
\end{align}
and
\begin{align}
2\nabla\Phi\cdot \v \Pi_A &= -2\Phi\Div\v \Pi_A =
2\Phi\ddot{\Phi}=2\Phi(\nabla^2\Phi+\varrho) \nonumber\\
&= -2\left(\nabla\Phi\right)^2+2\varrho\Phi
\end{align}
where integration by parts has been used liberally, since these are
integrands embedded in an outer integral, namely, the actual
Hamiltonian is related to the Hamiltonian density by an integral.
Using the above, then
\begin{align}
\mathscr{H}_{F,0}=\v E^2+\v B^2
=\mathscr{H}_{A,0}-\mathscr{H}_{\Phi,0}+2\varrho\Phi
\end{align}
The above is important for the derivation of \eqref{dch:eq8}.

\section{Laplace Transform Approach for Quantum Dissipation}

Defining the Laplace transform of $a(t)$ as $A(s)$, one deduce from
\eqref{qd:eq8} and \eqref{qd:eq9} that
\begin{align}\label{a1:eq7}
\hat A(s)\left( s+ i\omega_0 +
\sum_j\frac{\gamma_j^2}{s+i\omega_j}\right)=\hat a(0)
\end{align}

Assuming that there are infinitely many harmonic oscillators in the
bath, then the above summation can be replaced by an integral when
the number of modes is infinitely large, and the spacing between
their frequencies becomes infinitesimally small.  Namely,
\begin{align}
I(s)=\sum_j \frac{\gamma_j^2}{s+i\omega_j}&\Rightarrow
\frac1{2\pi}\int_{-\infty}^{\infty} d\omega
\frac{\eta(\omega)}{s+i\omega}\nonumber\\
&= \frac1{i2\pi}\int_{-\infty}^{\infty}
d\omega \frac{\eta(\omega)}{\omega-is}
\end{align}
where $\gamma_j^2\approx \frac1{2\pi}\eta(\omega)\Delta\omega$ and
$\omega=\omega_j=j\Delta\omega$. A pole is located at $\omega=is$.
But the radius of convergence (ROC) on the complex $s$ plane is for
$\Re e (s)
>0$. Therefore, for $s$ in the ROC, the pole in the complex $\omega$
is above the real $\omega$ axis.

 By assuming that
$\eta(\omega)\rightarrow 0,\quad |\omega|\rightarrow\infty$ and that
$\eta(\omega)$ is analytic, then by invoking Cauchy's theorem and
Jordan's lemma \cite{Hildebrand}, the above integral can be
evaluated in closed form yielding
\begin{align}
I(s)= \eta(is)
\end{align}
Since $\eta(\omega)$ is real when $\omega$ is real, $\eta(is)$ is
approximately real when $is$ is close to the real axis. But,
$\eta(s)$ is weakly dependent on $s$, and can be assumed to be
independent of frequency $\omega$. Then the pole of the system
described by \eqref{a1:eq7} is given by
\begin{align}
s=-i\omega_0-\eta
\end{align}
The above pole corresponds to a dissipative mode representing
quantum loss.

If the noise bath oscillators are not set to zero at $t=0$, then the
pertinent equation for the initial value problem becomes
\begin{align}
\hat A(s)\left( s+ i\omega_0 +
\sum_j\frac{\gamma^2}{s+i\omega_j}\right)=\hat
a(0)-i\sum_j\frac{\gamma_j\hat b_j(0)}{s+i\omega_j}
\end{align}
The summation on the left-hand side can again be approximated as
before to arrive at
\begin{align}
\hat A(s)\left( s+ i\omega_0 + \eta\right)=\hat
a(0)-i\sum_j\frac{\gamma_j \hat b_j(0)}{s+i\omega_j}
\end{align}
It is not possible to find a simple approximation to the summation
on the \RHS\ since $\hat b_j(0)$ is a random variable in the noise
bath. So it is left unchanged.  The above equation can be transformed
back to the time domain to yield
\begin{align}
\dot{\hat a}=-i\omega_0\hat a -\eta\hat  a -i\sum_j \gamma_j \hat
b_j(0)e^{-i\omega_j t}
\end{align}
The above is similar to \eqref{qd:eq17}.

\section{Proof of Preservation of Commutator}\label{app:3}

The proof of the preservation of the quantum commutator has been
lucidly presented by Tan in \cite{TAN}.  But since it is in Chinese,
it is reproduced here.
 By using the product rule for differentiation, one gets
\begin{align}\label{B1}
\frac{d}{dt} \left[\hat a,\hat a^\dag \right]=\left[ \hat a,
\frac{d}{dt} \hat a^\dag\right]+ \left[ \frac{d}{dt}\hat a, \hat
a^\dag\right]
\end{align}
From \eqref{pi:eq2} and \eqref{pi:eq3}, one gets
\begin{align}\label{B2}
\frac{d}{dt} \hat a^\dag(t)&=i\omega \hat a^\dag (t) -\eta \hat
a^\dag (t) + \hat F^\dag(t),\nonumber\\
\frac{d}{dt} \hat a(t)&=-i\omega
\hat a(t) -\eta \hat a (t) + \hat F(t)
\end{align}
From \eqref{B1}
\begin{align}\label{B3}
\frac{d}{dt} \left[\hat a,\hat a^\dag\right]=i\omega_0\left[ \hat a,
\hat a^\dag\right]-\eta \left[ \hat a, \hat a^\dag\right]+\left[\hat
a,\hat
F^\dag\right]\notag\\
-i\omega_0\left[ \hat a, \hat a^\dag\right]-\eta \left[ \hat a, \hat a^\dag\right]+\left[\hat F,\hat a^\dag\right]\notag\\
=-2\eta \left[ \hat a, \hat a^\dag\right]+\left[\hat a,\hat
F^\dag\right]+\left[\hat F,\hat a^\dag\right]
\end{align}
Integrating the above yields
\begin{align}
\hat a^\dag (t)=\hat a^\dag(0)e^{i\omega_0 t-\eta t}+\int_0^t d\tau
e^{(i\omega_0 -\eta )(t-\tau)}\hat F^\dag(\tau)\\
\hat a (t)=\hat a(0)e^{-i\omega_0 t-\eta t}+\int_0^t d\tau
e^{(-i\omega_0 -\eta )(t-\tau)}\hat F(\tau)
\end{align}
The commutators on the \RHS\ need to be evaluated, yielding
\begin{align}
\left[\hat F(t),\hat a^\dag (t)\right]&=\left[\hat F(t),\hat
a^\dag(0)e^{i\omega_0 t-\eta t}\right]\nonumber\\
&+\left[\hat F(t),\int_0^t d\tau\,
e^{(i\omega_0 -\eta )(t-\tau)}\hat F^\dag(\tau)\right]\\
\left[\hat a(t), \hat F^\dag(t) \right]&=\left[\hat a(0)e^{-i\omega_0
t-\eta t},\hat F^\dag(t)\right]\nonumber\\
&+\left[\int_0^t d\tau e^{(-i\omega_0
-\eta )(t-\tau)}\hat F(\tau),\hat F^\dag(t)\right]
\end{align}
Using the definition for the commutator as given in \eqref{pi:eq4},
the first term on the \RHS\ of the above can be shown to be zero.
Then it can be shown that the above becomes
\begin{align}
\left[\hat F(t),\hat a^\dag (t)\right]=\int_0^t d\tau e^{(i\omega_0
-\eta )(t-\tau)}\left[\hat F(t),\hat F^\dag(\tau)\right]
\end{align}
It is more prudent to take the ensemble average of the above, as the
noise bath can consist of time-varying dipoles, e.g., in Brownian
motion.   Then
\begin{align}
\left\langle\left[\hat F(t),\hat a^\dag
(t)\right]\right\rangle=\int_0^t d\tau e^{(i\omega_0 -\eta
)(t-\tau)}\left\langle\left[\hat F(t),\hat
F^\dag(\tau)\right]\right\rangle
\end{align}
From \eqref{pi:eq5} that
\begin{align}
\left\langle\left[\hat F(t),\hat F^\dag(t')\right]\right\rangle=\eta
\delta(t-t')
\end{align}
then
\begin{align}
\left\langle\left[\hat F(t),\hat a^\dag (t)\right]\right\rangle=
\left\langle\left[\hat a(t), \hat F^\dag(t)\right]\right\rangle=\eta
\end{align}
Finally, from \eqref{B3}, after taking ensemble average.
\begin{align}
\frac{d}{dt}\left\langle\left[\hat a(t),\hat
a^\dag(t)\right]\right\rangle=2\eta\left(1-\left\langle\left[\hat
a(t),\hat a^\dag(t)\right]\right\rangle\right)
\end{align}
The above implies that
\begin{align}
\left\langle\left[\hat a(t),\hat a^\dag(t)\right]\right\rangle=1
\end{align}
or that the commutator is preserved in the ensemble average sense..


\bibliography{basename of .bib file}

\begin{thebibliography}{100}
\bibitem{Explanation} In order to present in this paper a formulation of dissipative quantum electromagnetic systems with the least amount of concepts foreign to the classical electromagneticist, we have elected not to adopt the standard Coulomb gauge treatment and avoid using the Lagrangian formulation. In the Coulomb gauge, all redundancies in the electromagnetic system is removed. Hence, quantization can be done without any special treatment of the absent longitudinal vector potential and scalar potential (non-Coulomb potential part). Admittedly, in the Lorenz gauge used in this paper, one has to employ some post processing to correctly interpret the quantized $\hat{\mathbf{A}}_{\parallel}$ and $\hat{\Phi}$. However, the transverse and longitudinal dynamics in the field-matter system considered here completely decouples. It can be shown that $\mathbf{A}_{\parallel}$ and $\Phi$ evolves independent of the other dynamical variables. This is true for this system both classically and quantum mechanically. As such, the difference between using the Coulomb and Lorenz gauge is only formal. We will demonstrate the details of the decoupling as well as the necessary post processing in a future publication. The standard treatment can be found in classical texts such as \cite{Tannoudji,Scully&Zubairy,Garrison&Chiao,Grunner&Welsch}, and have been applied to the dissipative electromagnetic system in the literature \cite{Huttner&Barnett,Philbin}.
\bibitem{Footnote1}
It is to be noted that the coupling to the
Drude-Lorentz-Sommerfeld oscillator to obtain a classical dispersion
model for the medium has also been used in \cite{Raman&Fan}.
\bibitem{Footnote2}
In this unit, the velocity of light is one, and it
can be achieved by either redefining the unit for time or length
such that the velocity of light is one unit length per unit time.
Also, the units for $\v E$, $\v H$, and $\v P$ can be redefined such
that energy density is proportional to $\v E^2$, $\v H^2$, and $\v
P^2$.  Here, the unit for length is redefined so that one unit
length is $\approx 3\times 10^8$ m for one second.  Alternatively,
one can take one unit length to be $\approx 0.3$ m for one
nano-second.
\bibitem{Footnote4}
The raison d'etre for this will be given in our future work.
\bibitem{Footnote3}
This equation is found in standard quantum theory text,
and also explained in \cite{QEM1}.
\bibitem{Footnote5}
Notice the sign difference for the canonical commutator of the scalar potential. This sign reversal is needed due to the negative contribution of the scalar potential in the Hamiltonian.
\bibitem{Footnote7}
The bath oscillators are mathematically similar to the Lorentz oscillator.
\end{thebibliography}

\end{document}